\documentclass[fleqn,usenatbib]{mnras}

\usepackage{booktabs}
\usepackage{tablefootnote}
\usepackage{subfloat}

\usepackage[T1]{fontenc}

\DeclareRobustCommand{\VAN}[3]{#2}
\let\VANthebibliography\thebibliography
\def\thebibliography{\DeclareRobustCommand{\VAN}[3]{##3}\VANthebibliography}

\usepackage{graphicx}	
\usepackage{amsmath}	
\usepackage{amssymb}	
\newcommand{\angstrom}{\text{\normalfont\AA}}
\usepackage{newtxtext,newtxmath}
\usepackage{tablefootnote}
\usepackage{pdflscape}

\title[SEAMBH]{X-ray properties of reverberation-mapped AGNs with super-Eddington accreting massive black holes}

\author[J. Maithil et al.]{Jaya Maithil,$^{1}$\thanks{E-mail: jaya.maithil@cfa.harvard.edu}
Michael S. Brotherton,$^{2}$ Ohad Shemmer,$^{3}$ Bin Luo,$^{4,5}$ Pu Du,$^{6}$ Jian-Min Wang,$^{6,7,8}$ Hu Chen,$^6$
\newauthor{ Sarah C. Gallagher,$^9$  Yan-Rong Li,$^6$ \& Rodrigo S. Nemmen. $^{10}$}
\\
$^{1}$Center for Astrophysics| Harvard \& Smithsonian, 60 Cambridge Street, Cambridge, MA 02138, USA \\
$^{2}$Department of Physics and Astronomy, University of Wyoming, Laramie, WY 82071, USA \\
$^{3}$Department of Physics, University of North Texas, Denton, TX 76203, USA \\
$^{4}$School of Astronomy and Space Science, Nanjing University, Nanjing, Jiangsu 210093, China \\
$^{5}$Key Laboratory of Modern Astronomy and Astrophysics
(Nanjing University), Ministry of Education, Nanjing 210093, China \\
$^{6}$Key Laboratory for Particle Astrophysics, Institute of High Energy Physics, Chinese Academy of Sciences, 19B Yuquan Road, Beijing 100049,\\ People’s Republic of China \\
$^{7}$National Astronomical Observatories of China, Chinese Academy of Sciences, 20A Datun Road, Beijing 100020, People’s Republic of China \\
$^{8}$School of Astronomy and Space Science, University of Chinese Academy of Sciences, 19A Yuquan Road, Beijing 100049, People’s Republic of China \\
$^{9}$Department of Physics \& Astronomy, University of Western Ontario, London, ON N6C 1T7, Canada \\
$^{10}$Instituto de Astronomia, Geofísica e Ciências Atmosféricas, Universidade de Sao Paulo,  05508-090 São Paulo, SP, Brazil}

\date{Accepted 2024 January 3. Received 2023 December 14; in original form 2023 April 25}

\pubyear{2024}

\begin{document}
\label{firstpage}
\pagerange{\pageref{firstpage}--\pageref{lastpage}}
\maketitle

\begin{abstract}
The X-ray properties of Active Galactic Nuclei (AGNs) depend on their underlying physical parameters, particularly the accretion rate.  We identified eight reverberation-mapped AGNs with some of the largest known accretion rates without high-quality X-ray data.  We obtained new Chandra ACIS-S X-ray observations and nearly simultaneous optical spectrophotometry to investigate the properties of these AGNs with extreme super-Eddington accreting black holes (SEAMBHs).  We combined our new X-ray measurements with those of other reverberation-mapped AGNs, which have the best-determined masses and accretion rates.  The trend of the steepening of the spectral slope between X-ray and optical-UV, $\alpha_{\rm ox}$, with increasing optical-UV luminosity, $L_{\rm 2500\angstrom}$, holds true for even the most extreme SEAMBHs. One of our new SEAMBHs appears X-ray weak for its luminosity, perhaps due to absorption associated with orientation effects involving a slim disk thought to be present in highly accreting systems. The correlation of the $\rm 2-8~ keV$ X-ray photon index with the accretion rate also holds for the extreme SEAMBHs, which show some of the largest photon indices reported for AGNs.
\end{abstract}



\section{Introduction}\label{intro}
Active Galactic Nuclei (AGNs) emit radiation across the entire electromagnetic spectrum due to various physical processes ultimately powered by the gravitational potential energy of the accreting supermassive black hole \citep{Salpeter1964,Pringle1973,Ho2008}. The accreting matter forms a disk around the central supermassive black hole, which glows brightly in optical, UV, and soft X-rays. The spectral energy distribution of AGNs peaks in the UV emitted from the inner accretion disk. The Compton upscattering of optical-UV disk photons by the hot electrons in a region close \citep[observations indicate three to few tens of gravitational radii,][]{SunyaevTruemper1979, McHardy+2005, Wilkins+2016, Chartas+2016} to the black hole, called the corona, produces hard X-rays \citep[e.g.,][]{Haardt+Maraschi1991,Haardt+Maraschi1993, Nakamura+1993, Kawaguchi+2001}. Above 2 keV (in the range 2-100 keV), the X-ray spectrum is dominated by a single power-law continuum, $N(E)\propto E^{-\Gamma}$, with a high-energy cut-off \citep{SunyaevTitarchuk1980}. Studies have found the cut-off energy is typically tens of keV or higher \citep{Ricci+2017, Molina+2019, Tortosa+2023}. The photon index ($\Gamma$) acts as a tool to infer the energy distribution of the electrons in the corona. The two-point spectral index between 2500 $\angstrom$  and 2 keV \citep[$\alpha_{\rm ox}$,][]{Tananbaum+1979, Just+2007}{}{} indicates the relative strength of the energy emitted by the disk versus the corona.

Past studies have examined the correlation between the X-ray-to-optical properties of quasars with various observed and fundamental properties. The photon index and $\alpha_{\rm ox}$ generally show no variation with redshift \citep[e.g.,][]{Page+2005,Shemmer+2005, Shemmer+2006, Steffen2006,Just+2007}.  However, $\alpha_{\rm ox}$ exhibits a strong anti-correlation with optical-UV luminosity at 2500 $\angstrom$ \citep{Steffen2006, Just+2007,Timlin+2020}. \cite{Steffen2006} find a 13.6$\sigma$ correlation between $\alpha_{\rm ox}$ and $L_{2500\angstrom}$ using a sample of 333 AGNs with redshifts up to $z\sim 6$ and spanning over five orders of magnitude in $L_{2500\angstrom}$ and four orders of magnitude in X-ray luminosity.  The $\alpha_{\rm ox}$-$L_{2500\angstrom}$ correlation is a by-product of the non-linear correlation between the luminosities at 2 keV and 2500 $\angstrom$ \citep{Vignali+2003,Strateva+2005,Steffen2006, Just+2007, Lusso+2010,Young+2010}. It indicates AGNs with higher optical brightness emit relatively fewer X-rays than more optically faint AGNs \citep{Lusso+Risaliti2016}. Through careful sample control, \cite{Lusso+Risaliti2016} tightened the $L_{2500\angstrom}$-$L_{2 \rm~ keV}$ relation, reducing the dispersion from $\sim$0.35–0.4 dex (from previous studies) to $\sim$0.21–0.24 dex seen in four orders of magnitude in luminosity. Such a tight correlation indicates that the energy emitted by the accretion disk and corona are inextricably linked. We also note that $\alpha_{\rm ox}$ shows weak to no correlation with the Eddington ratio \citep[e.g.,][]{Young+2010}.

A strong anti-correlation between $\Gamma$ (both soft and hard photon index) and full-width half maxima of H$\beta$ emitted from the broad emission-line region \citep{Wang+1996,Laor+1997, Brandt+1997, Wang+2004} indicates the dependence of X-ray properties on the fundamental properties of black holes like mass or accretion rate. This anti-correlation is likely a secondary effect of a more intrinsic correlation between hard $\Gamma$ and the Eddington ratio ($L_{\rm Bol}/L_{\rm Edd}$), a normalized accretion rate parameter \citep{Shemmer+2006,Shemmer+2008, Brightman+2013, Trakhtenbrot+2017, Tortosa+2023}. As the accretion rate increases, the accretion disk heats up, glowing in soft X-rays and increasing the Compton cooling of the corona--resulting in a steepening and softening of the X-ray spectra  \citep{Shemmer+2006,Shemmer+2008}. Therefore, X-rays are critical for probing the accretion process in the vicinity of black holes.

The correlation between the $\rm 2-8~ keV$ photon index and Eddington ratio holds over a range of quasar luminosity and redshift. E.g., \cite{Shemmer+2008} combine the \cite{Shemmer+2006} sample of 25 moderate luminosity ($43 < {\rm log}~ \nu L_{\nu}(5100\angstrom) [\rm erg~ s^{-1}] \leq 46$) and five high luminosity ($46 < {\rm log}~ \nu L_{\nu}(5100\angstrom) [\rm erg~ s^{-1}] \leq 48$) type-1 radio-quiet quasars at $z<0.5$ and $z\sim2$, respectively, with five high luminosity quasars at $z=1.3-3.2$. These 35 quasars have high-quality optical spectra to derive black hole mass estimates based on the
H-beta line and high-quality X-ray spectra to measure $\Gamma$. Although the
sample was not complete or necessarily representative, their results established
a significant correlation between $\Gamma_{\rm 2-8~ keV}$ and $L_{\rm Bol}/L_{\rm Edd}$ over four orders of magnitude in quasar luminosity.  \cite{Risaliti+2009} studied $\sim$400 AGNs with good-quality optical spectra from SDSS and XMM-Newton X-ray observations from the SDSS/XMM-Newton quasar survey \citep{Young+2009} spanning three orders of magnitude in optical luminosity. Their results show a highly significant correlation between $\Gamma_{\rm 2-8~ keV}$ and $L_{\rm Bol}/L_{\rm Edd}$. \cite{Risaliti+2009} also find that the strength and significance of the $\Gamma_{\rm 2-8~ keV}$-$L_{\rm Bol}/L_{\rm Edd}$ correlation decreases going from H$\beta$ to Mg {\sc ii} to no correlation when C{\sc iv}-based mass is used to calculate the Eddington ratio. \cite{Brightman+2013} also confirmed the $\Gamma_{\rm 2-8~ keV}$-$L_{\rm Bol}/L_{\rm Edd}$ correlation using a sample of 69 type-1 radio-quiet AGNs with redshifts up to $\sim 2.1$ with X-ray spectra from the Chandra Deep Field-South survey \citep{Lehmer+2005} and black hole mass estimates based on H$\alpha$ and Mg {\sc ii} measurements from the Cosmic Evolution Survey \citep{Cappelluti+2009, Elvis+2009}. Using 71 type-1 AGNs selected from the XMM-Newton Bright Serendipitous survey, \cite{Fanali+2013} found a significant correlation between $\Gamma_{0.5-10~ \rm keV}$ and  $L_{\rm Bol}/L_{\rm Edd}$ after removing the dependence on redshift. They found a less significant correlation with $\Gamma_{\rm 2-8~ keV}$ due to a larger error in determining $\Gamma_{\rm 2-8~ keV}$, a factor of 2.5 larger compared to the error in $\Gamma_{0.5-10~ \rm keV}$. 

The correlations like $\Gamma_{\rm 2-8~ keV}$-$L_{\rm Bol}/L_{\rm Edd}$ and $\alpha_{\rm ox}-L_{2500\angstrom}$ are evident for the sub-Eddington accreting ($L_{\rm Bol}/L_{\rm Edd} \ll 1$) sources. Quasars with low Eddington ratios presumably have an optically thick and geometrically thin accretion disk. In the standard \cite{Shakura+Sunyaev1973} thin disk model, the spin of the black hole defines the innermost stable orbit that dictates the radiation efficiency ($\eta$). For a retrograde spin, $\eta$ = 0.038, whereas for a maximally spinning black hole, $\eta$ = 0.32. But in the case of a higher accretion rate, the accretion disk becomes geometrically thick or slim \citep{Abramowicz+1988,Laor+Netzer1989,Wang+1999, Wang+2013}. Due to effects like photon trapping and advection-dominated energy transport, the slim disk has significantly smaller radiation efficiency that may be independent of spin \citep{Wang+1999,Mineshige+2000,Wang+2013}. The spectral energy distribution (SED) of a slim accretion disk is likely different from a thin disk with a cut-off at higher energies. Although a comparison of IR-optical-UV-X-ray SEDs of super- and sub-Eddington RM AGNs show no anomalous torus emission or a heightened ionizing continuum predicted for slim disk systems \citep{Castello+2016,Castello+2017}. Super-Eddington accreting AGNs exhibit optical-UV SEDs that align well with the standard thin disk model. Any indications of a slim disk, may be present in the extreme ultra-violet (EUV) where data availability is limited \citep{Castello+2016, Kubota_Done2019}.

\cite{Wang+2014} suggested that high accreting objects likely have slim accretion disks and coined the acronym Super-Eddington accreting massive black holes (SEAMBHs). They are characterized by large Eddington ratios ($L_{\rm Bol}/L_{\rm Edd}>0.3$). 
SEAMBHs deviate from the traditional radius-luminosity relationships established for the low-accreting AGNs of the form $R \propto L^{\sim 0.5}$ \citep[e.g.,][]{Kaspi+2000, Bentz+2013}. 
A dedicated RM campaign by \cite{Du+2014,Du+2015,Du+2016, Du+2018} tested the R-L relationship for the most highly accreting AGNs. They selected SEAMBH candidates based on a dimensionless accretion rate estimator, $\dot{\mathscr{M}}$, such that a $\dot{\mathscr{M}}\geq 3$ implies a SEAMBH. In context of thin-disks, the Eddington ratio is related to $\dot{\mathscr{M}}$ by the equation $L_{\rm Bol}/L_{\rm Edd} = \eta \dot{\mathscr{M}}$ \citep[][]{Du+2014, Du+2015}{}{}. Here, $\eta$ is the mass-to-radiation conversion efficiency, a parameter dependent on the black hole spin. \cite{Du+2018} shows that the highest accretion rate AGNs have systematically shorter time lags, a factor of 3-8 times smaller than predicted by the canonical R-L relationship for the sub-Eddington accreting AGNs. Slim accretion disks in SEAMBHs are perhaps responsible for the shortened time lags of the broad-line region, due to anisotropic emission from the ionizing source \citep{Wang+2014, Du+Wang2019}. The canonical single-epoch black hole masses of highly accreting quasars are overestimated, on average, by a factor of two and the accretion rate parameters  $L_{\rm Bol}/L_{\rm Edd}$ and $\dot{\mathscr{M}}$ consequently underestimated \citep[][]{Maithil+2022}. SEAMBHs are not so common at low redshift but their fraction is expected to be higher in the early universe \citep{Kelly+Shen2013}. Discoveries of billion solar mass black holes at $z>6$, when the universe was less than a billion years old, are suggestive of super-Eddington accretion (e.g., review by \cite{Valiante+2017} and references therein). Thus, SEAMBHs are probes to understand the accretion process and disk-corona connection of the high-redshift massive highly accreting AGNs.

Recently, \cite{Liu+2021} probed the disk-corona connection using a sample of 26 sub-Eddington and 21 super-Eddington accreting AGNs. They used the most accurate reverberation-mapped black hole masses eliminating a major source of uncertainty in the $\Gamma_{\rm 2-8~ keV}$-$L_{\rm Bol}/L_{\rm Edd}$ correlations coming from the black hole mass estimates used to derive the Eddington ratio.  Other intrinsic differences between AGNs like black hole spin, orientation, and optical depth in the corona likely also contribute to uncertainties in such relations. Most of the past studies were limited to sub-Eddington sources and used the single-epoch black hole mass scaling relations for broad-emission lines like H$\beta$, H$\alpha$, and Mg {\sc ii} based on the traditional R-L relationships. Some studies looked at X-ray and optical-UV properties of super-Eddington accreting sources but they did not perform a comparative investigation with the sub-Eddington sources \citep[e.g.,][]{Ai+2011,Kamizasa+2012}. In this paper, we extend the investigations of Liu et al. (2021) by adding 13 extreme SEAMBHs from the SEAMBH-RM campaign that show the highest accretion rate (log $\dot{\mathscr{M}}>1.5$) and the largest deviation in the broad-line region (BLR) sizes from the canonical R-L relationship. We present the Chandra X-ray data of 8 SEAMBHs for the first time with simultaneous optical-UV spectra avoiding uncertainties in the derivation of $\alpha_{\rm ox}$ due to variability. Our work tests whether the extreme SEAMBHs follow the trends seen in sub- and super-Eddington accreting AGNs. In Section \ref{section2}, we describe our sample and summarize the sample selection of Liu et al (2021). Section \ref{section3} presents the data reduction and analysis of our new Chandra observation and the optical-UV spectra. This is followed by an explanation of measured and derived quantities in Section \ref{section4}. Section \ref{section5} presents the key results and Section \ref{section6} presents the discussion and conclusion. Throughout the paper, we adopt a cosmology with $H_0 = 70~ {\rm km s^{-1} Mpc^{-1}}, \Omega_{\Lambda} = 0.7$ and $\Omega_m = 0.3$.

\begin{table*}
  \caption{Sample properties and observation log of the eight extreme SEAMBHs selected for Chandra observation and one Archived SEAMBH.}
    \label{tab:table1}
\makebox[\textwidth][c]{
    \begin{tabular}{ccccccccccc}
\toprule
    &   & &   &  & \multicolumn{3}{c}{Chandra X-ray} & \multicolumn{2}{c}{Optical} \\
\cmidrule(lr){6-8}\cmidrule(lr){9-10}
Object  &   z   & $N_{\rm H}$   &   log $M_{\rm BH}$   &   log $\dot{\mathscr{M}}$ &   ObsID   & Date  &   Exposure time   &   Observatory &   Date\\
  &   &  ($10^{20} \rm cm^{-2}$) & ($M_{\sun}$)  &    &     &   &  (s)   &    &    \\   \midrule
IRAS04416+1215				&	0.09	& 15.45	&	6.78	&	2.63	&	21510	&	2018-10-23	&	3466.68	&	Lijiang & 2018-10-25	\\
SDSSJ074352.02+271239.5	&	0.25	& 4.24	&	7.93	&	1.69	&	21511	&	2018-10-27	&	4551.67	&	CAHA & 2018-11-04	\\
SDSSJ080131.58+354436.4	&	0.18	& 4.90	&	6.51	&	2.43	&	21512	&	2018-12-16	&	7350.62	&	Lijiang & 2018-12-19	\\
SDSSJ083553.46+055317.1	&	0.20	&	3.19	&	6.87	&	2.41	&	21513	&	2019-01-15	&	8558.59	&	CAHA & 2019-01-17	\\
SDSSJ084533.28+474934.5	&	0.30		&  2.96 &	6.76	&	2.76	&	21514	&	2019-01-12	&	17319.36	&	Lijiang & 2019-01-18	\\
SDSSJ093302.68+385228.0	&	0.18	& 1.44	&	7.08	&	1.79	&	21515	&	2019-01-19	&	10897.60	&	Lijiang & 2019-01-24	\\
SDSSJ100402.61+285535.3	&	0.33	&  2.09	&	7.44	&	2.89	&	21516	&	2018-11-18	&	6311.02	&	Lijiang & 2018-11-20	\\
SDSSJ101000.68+300321.5	&	0.26	&  2.27	&	7.46	&	1.70	&	21517	&	2019-01-23	&	9864.07	&	Lijiang	& 2019-01-25\\
\midrule
SDSSJ075051.72+245409.3$^\dagger$ & 0.40 & 5.32 & 7.67 & 2.14 & 24723 & 2021-03-29 & 3387.21 & ... & ... \\
\bottomrule
\end{tabular}%
}

\footnotesize{Notes. - Redshift, black hole mass, and dimensionless accretion rate measurements are from \cite{Du+Wang2019}. Galactic neutral hydrogen column density ($N_{\rm H}$) is from \cite{DickeyLockman1990}}. Chandra exposure time (or LIVETIME) is the ONTIME corrected for average dead time corrections.\\ $^\dagger$This target was observed by Chandra (PI: Gordon Garmire)  which is analyzed in this paper and is part of the Archived SEAMBHs sub-sample.

\end{table*}

\section{Sample}\label{section2}

The reverberation mapping of SEAMBHs by \cite{Du+2018} leads to the identification of 14 SEAMBHs that have markedly smaller BLR sizes than expected from their luminosities and have extreme accretion rates, i.e., log $\dot{\mathscr{M}}>1.5$. Our core sample consists of these 14 `extreme SEAMBHs'. 

Eight of these extreme SEAMBHs did not have X-ray observations above 2 keV (called `Chandra SEAMBHs'). We proposed Chandra Advanced CCD Imaging Spectrometer (ACIS-S) observations of these eight SEAMBHs in Cycle 20. We used the timed exposure ACIS operation mode with the faint telemetry format. To mitigate potential pile-up issues, we used the 1/8 sub-array with only the S3 chip turned on. The spectral coverage of Chandra allows us to measure both the 2-8 keV X-ray photon index ($\Gamma_{2-8~ \rm keV}$) and, combined with optical-UV data, the $\alpha_{\rm ox}$. To minimize the effect of variability in $\alpha_{\rm ox}$ measurements, the Chandra observations were accompanied by ground-based optical-UV spectral observations as close in time as weather permitted using facilities like the Lijiang  2.4 m and the 2.2 m Calar Alto telescopes. We present the X-ray and optical-UV data reduction and analysis in the next section. Table \ref{tab:table1} provides the observation log of our Chandra targets. 

The remaining six extreme SEAMBHs have archival data from Chandra, XMM-Newton, and Swift. SDSSJ075051.72+245409.3 has archival ASIC-S Chandra observation from Cycle 21 (PI: Gordon Garmire, see Table \ref{tab:table1} for details). In the absence of simultaneous optical-UV observation for this target, we use $L_{2500\angstrom}$ measurement from \cite{Shen+2011}. SDSS J080101.41+184840.7, SDSS J093922.89+370943.9, IRAS F12397+3333 and PG 2130+099 have simultaneous X-ray observations from the XMM-Newton's European Photon Imaging Camera-PN and MOS detectors and optical-UV observations from the Optical Monitor. Lastly, Mrk 142 is observed simultaneously in the X-ray and the UV-Optical using the Swift observatory. Throughout this paper, we refer to these six extreme SEAMBHs as `Archived SEAMBHs'. Except for SDSSJ075051.72+245409.3, the other five extreme SEAMBHs are part of the Liu et al. (2021) sample and we refer the readers to their Section 2 for details of X-ray and optical-UV analysis. We process the X-ray data of SDSSJ075051.72+245409.3 along with the eight Chandra SEAMBHs and present the details in Section \ref{section3a}. 

\begin{figure}
   \includegraphics[width=\columnwidth]{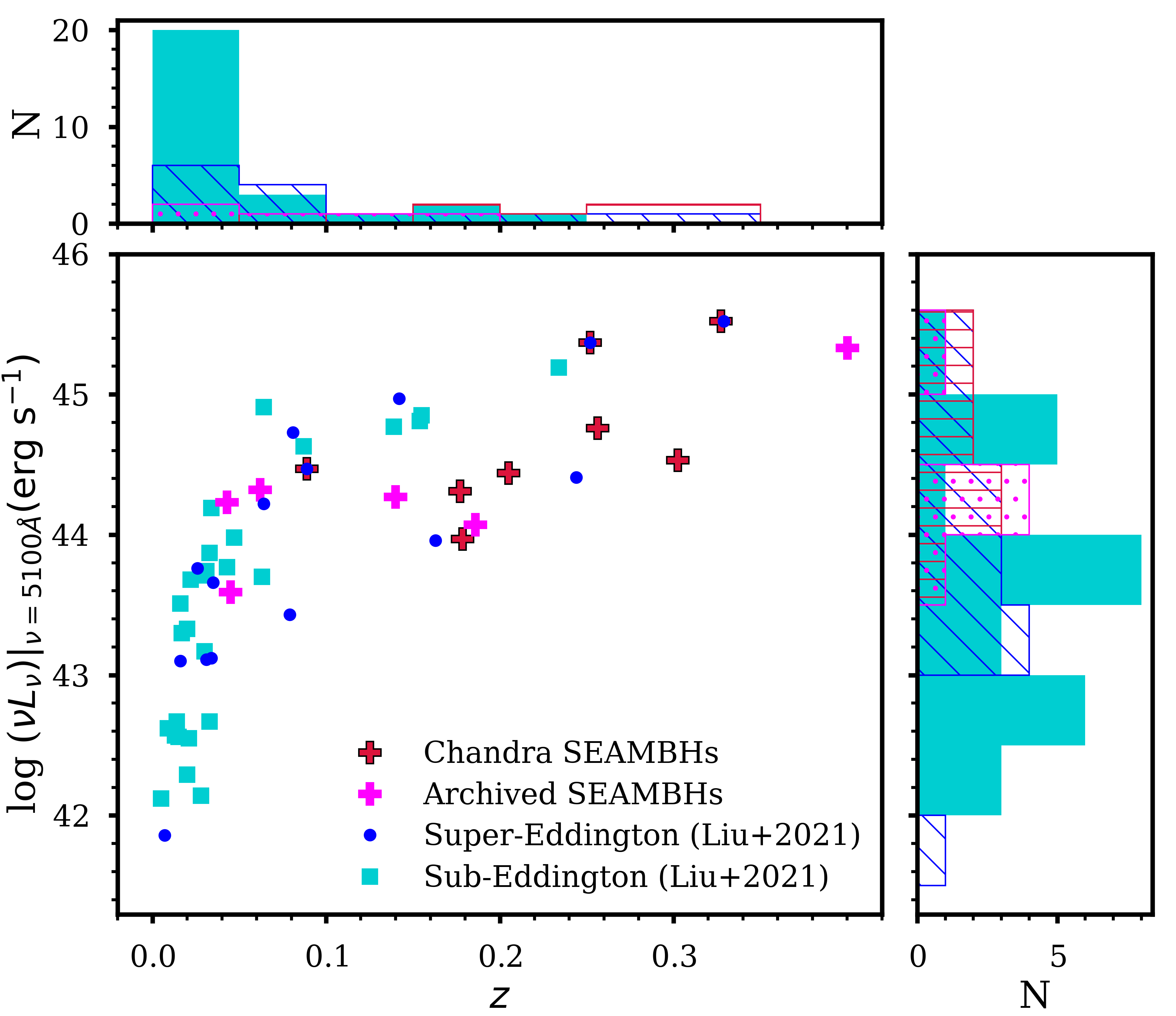}
   \caption{Monochromatic luminosity at 5100 $\angstrom$ as a function of redshift. The plot shows the redshift-luminosity space spanned by our sample.}
   \label{fig:Figure1}
\end{figure}

We use the remaining \cite{Liu+2021} sample of 16 super-Eddington ($0.47 < {\rm log}\dot{\mathscr{M}} < 1.5$) and 26 sub-Eddington accreting AGNs to complement our sample. \cite{Liu+2021} selected radio-quiet RM AGNs from \cite{Du+2015,Du+2016, Du+2018} that have good signal-to-noise archival X-ray data ($S/N>6$ in the rest-frame $>2$ keV band) and simultaneous optical-UV observations. They eliminated AGNs that are heavily absorbed in X-rays due to outflows or display broad-absorption-line features and reddening in their UV spectra. Three out of these 16 super-Eddington AGNs are part of our Chandra SEAMBHs. \cite{Liu+2021} present the archival Swift (IRAS 04416+1215, SDSS J074352.02+271239.5) and XMM-Newton (SDSS J100402.61+285535.3) data of these in-common targets. As X-rays can vary, we present the archival data of these three targets as part of the \cite{Liu+2021} super-Eddington sample and treat them as separate measurements.  Figure \ref{fig:Figure1} shows the $L_{5100\angstrom}-z$ plane for our 14 extreme SEAMBHs and the \cite{Liu+2021} sample, adding more data points at $z>0.16$. Our extreme SEAMBHs have log $L_{5100 \angstrom}>43.3$ placing them at the luminous end of the super-Eddington AGNs studied previously.

\begin{table*}
 \caption{Photon counts in broad, hard, and soft energy bands, and the hardness ratio.}
 \label{tab:table2}
 \begin{tabular}{lllllll}
 \hline
Object	&	Broad Band	&	Hard Band (H) &	Soft Band (S) &	Hardness Ratio	\\
    &   (0.35-8 keV)    &   (2-8 keV)   &   (0.35-2 keV) & (H-S)/(H+S) \\
 \hline
IRAS04416+1215	&	1291.71$\pm$35.98	&	474.08$\pm$21.80	&	817.63$\pm$28.62 & -0.27    \\
SDSSJ074352.02+271239.5	&	245.58$\pm$15.69	&	143.90$\pm$12.00	&	101.69$\pm$10.10  & 0.17  \\
SDSSJ080131.58+354436.4	&	1015.00$\pm$31.86	&	188.00$\pm$13.71	&	827.00$\pm$28.76	& -0.63\\
SDSSJ083553.46+055317.1	&	1356.97$\pm$36.88	&	322.64$\pm$18.01	&	1034.33$\pm$32.19	& -0.52\\
SDSSJ084533.28+474934.5	&	390.48$\pm$19.86	&	108.16$\pm$10.50	&	282.32$\pm$16.86 & -0.45	\\
SDSSJ093302.68+385228.0	&	452.87$\pm$21.34	&	111.48$\pm$10.64	&	341.39$\pm$18.50	& -0.51 \\
SDSSJ100402.61+285535.3	&	610.01$\pm$24.74	&	157.24$\pm$12.57	&	452.78$\pm$21.31 & -0.48	\\
SDSSJ101000.68+300321.5	&	72.69$\pm$8.68	&	49.31$\pm$7.16	&	23.38$\pm$4.91	&	0.36\\ \midrule
SDSSJ075051.72+245409.3 & 125.38$\pm$11.23 & 36.69$\pm$6.09 & 88.69$\pm$9.44 &  -0.41\\
 \hline
 \end{tabular}
\end{table*}
\section{Data reduction \& analysis}\label{section3}
\subsection{Chandra X-ray reduction and spectral analysis}\label{section3a}

This paper presents Chandra data made available after CXC's fifth reprocessing campaign. We analyzed data using Chandra Interactive Analysis of Observations software (CIAO) v4.13. The level 1 event files were reprocessed using the \texttt{chandra\_repro} script with CALDB v4.9.5 to create the level 2 event files and observation-specific bad pixel files. We extracted the net counts, i.e., source counts minus background counts from the level 2 event file. To extract source counts we used a circular aperture of 4\arcsec (2\arcsec for SDSS J074352.02+271239.5) at the source coordinate, whereas an annulus with an inner radius of 8\arcsec (5\arcsec) and an outer radius of 13\arcsec (10\arcsec) was used to extract the background counts. The source and background regions were saved for later use in spectral analysis. The net counts were obtained in three bands: Hard (2-8 keV), Soft (0.35-2 keV), and Broad (0.35-8 keV). Using these counts we calculated the hardness ratio (HR) defined as (Hard-Soft)/(Hard+Soft). Table \ref{tab:table2} lists the photon counts and hardness ratio. Six out of eight Chandra SEAMBHs show a negative hardness ratio implying soft spectra, whereas two (SDSS J074352.02+271239.5 and SDSS J101000.68+300321.5) show hard spectra. One Archived SEAMBH with Chandra data, SDSSJ075051.72+245409.3, presents a soft spectrum.

We used \texttt{specextract} to create source and background spectra and the instrument response files with point source aperture correction to unweighted auxiliary response files. The saved circular source aperture and annulus background regions were used to extract the spectra. We used \verb'Sherpa' to fit the spectral data.
The source model employed is a one-dimensional power-law and an XSpec photoelectric absorption model with the hydrogen column density frozen to the Galactic value \citep[][]{DickeyLockman1990} in the source direction (\texttt{xsphabs.abs1 * powlaw1d.p1}). 
The data were grouped to 5-30 counts per bin depending on the net count of the object. We used the Levenberg-Marquardt optimization method with W-statistic for fitting the spectra of all objects. The W-statistic is equivalent of the Cash statistic \citep[][]{Cash1979}, appropriate when using background-subtracted source spectra of low-count sources and also works well for high-count sources \citep[][]{Kaastra2017}. 
Figure \ref{fig:Figure2} shows the best-fit model for the observed-frame 0.35 to 8/(1+z) keV spectra. The best-fit parameter values for the photon index, $\Gamma_{\rm 0.35-8~ \rm keV}$, absorption-corrected rest-frame 0.35-8 keV flux, and the reduced statistics are presented in Table \ref{tab:table3}.

Next, we fit the observed-frame 2/(1+z) to 8/(1+z) keV X-ray band, where the underlying spectrum is mostly free of absorption or soft-excess emission \citep{Shemmer+2006, Shemmer+2008, Risaliti+2009, Brightman+2013}.
Six Chandra SEAMBHs and the Archived SEAMBH (SDSSJ075051.72+245409.3) show soft, steep spectra well fit by using a power-law model with Galactic absorption. SDSS J074352.02+271239.5 \& SDSS J101000.68+300321.5 appear as outliers and have much harder spectra than the other SEAMBHs. We fit their spectrum using a power-law model with both Galactic and intrinsic absorption (\texttt{xsphabs.abs1 * xszphabs.zabs1 * powlaw1d.p1}). SDSS J101000.68+300321.5 likely needs a more complex model, but the low photon count prevents us from uniquely doing so. For IRAS 04416+1215, we also adopted a high-energy cut-off for the power-law fixed to 44 keV from \cite{Tortosa+2023}. An F-test reveals that adding a high-energy cut-off doesn't improve the statistics significantly.
We tested the presence of intrinsic absorption for each source. Except for SDSS J074352.02+271239.5, an F-test confirms that adding intrinsic absorption component to the Galactic-absorbed power-law model does not improve the fit significantly.
Table \ref{tab:table3} reports the $2-8~ \rm keV$ photon index ($\Gamma_{\rm 2-8~ keV}$), the absorption-corrected rest-frame 2-8 keV flux, reduced statistics, and the monochromatic flux density at rest-frame energy of 2 keV from the best-fit models. It should be noted that the rest-frame $2-8~ \rm keV$ best-fit values are the key X-ray results that we will use in our analysis. We provide the literature values of the photon index for IRAS 04416+1215, SDSS J074352.02+271239.5, SDSS J100402.61+285535.3 and SDSSJ075051.72+245409.3 in the appendix \ref{appendix}.

To assess absorption or excess in the soft ($<2~ \rm keV$) X-ray band, we extrapolate the $\rm 2-8~ keV$ X-ray power-law model to the observed-frame 0.35 to 2/(1+z) keV energies. Nearly half of our targets exhibit soft excesses, with exceptions being SDSS J074352.02+271239.5, SDSS J084533.28+474934.5, SDSS J093302.68+385228.0 and SDSS J075051.72+245409.3. We tested three models to better describe their $<2~ \rm keV$ X-ray emission spectrum. Model 1 comprises a simple power-law with Galactic absorption. Model 2 includes a thermal Comptonization (\texttt{xscomptt}) component with a fixed underlying power-law using parameters from the $\rm 2-8~ keV$ X-ray analysis presented in Table \ref{tab:table3}. Model 3 includes an additional partial covering ionized absorption component (\texttt{ZXIPCF}) into the fitting to account for weak absorption in the soft ($<2~ \rm keV$) X-rays. For SDSSJ074352.02+271239.5, SDSSJ084533.28+474934.5, J093302.68+385228.0 and SDSS J075051.72+245409.3, the F-test indicates that fitting the $<2~ \rm keV$ band with a Galactic absorbed power-law model yields results similar to extending the $\rm 2-8~ keV$ best-fit to softer energies. Only for IRAS 04416+1215, SDSSJ080131.58+354436.4, SDSSJ083553.46+055317.1, SDSSJ100402.61+285535.3, and SDSSJ101000.68+300321.5, does Model 1 emerge as the best fit. Table \ref{tab:table4} gives the 0.35-2/(1+z) keV photon index, absorption-corrected rest-frame 0.35-2 keV flux and statistics of the best-fit model.   

\begin{subfigures}\label{fig:Figure2}
\begin{figure*}
  \includegraphics[width=2\columnwidth]{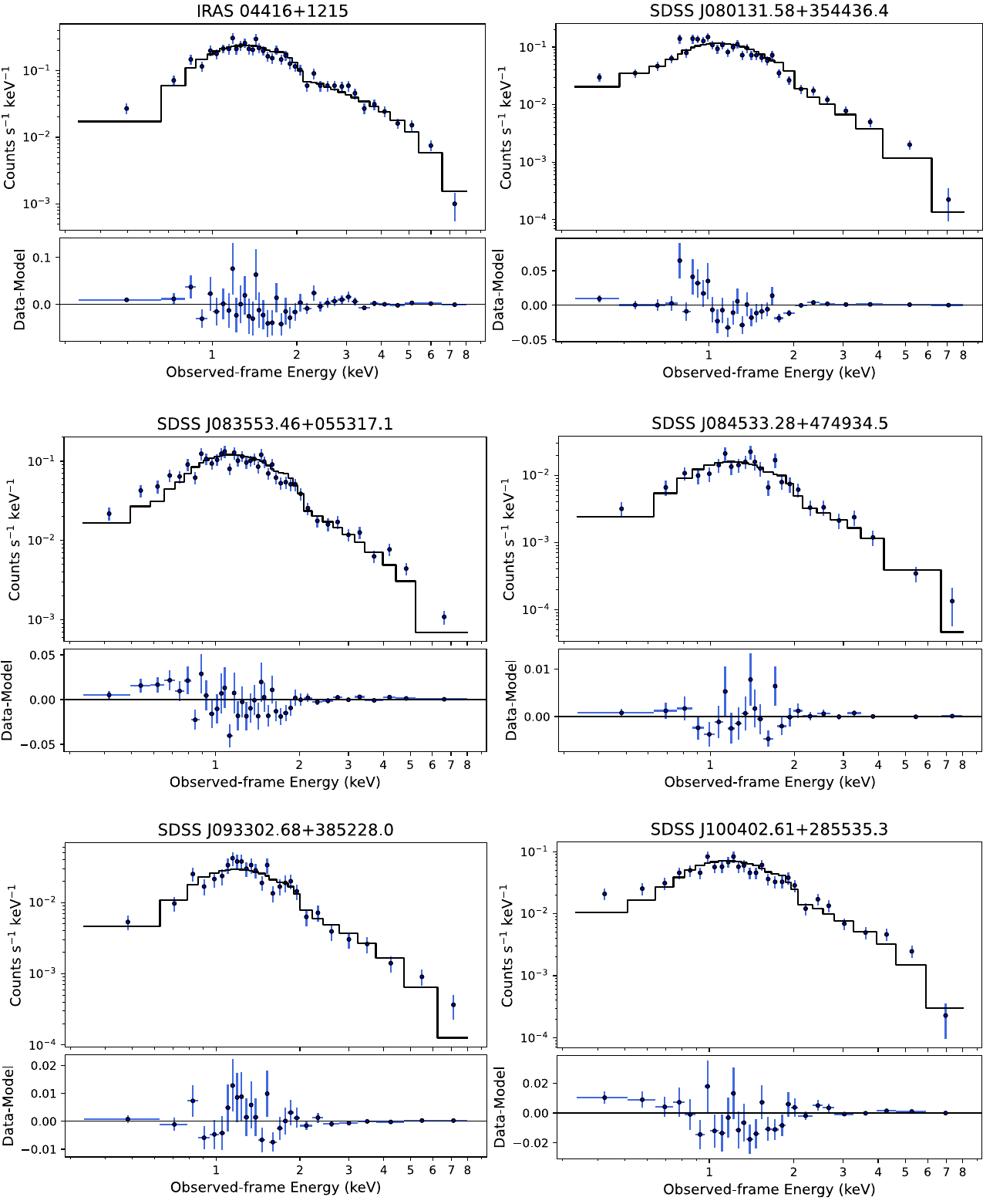}
  \caption{\label{first} Power-law fit with Galactic absorption to the observed-frame 0.35-8/(1+z) keV spectra of six Chandra SEAMBHs.} 
  \label{fig:Figure2a}
\end{figure*}
\begin{figure*}
  \includegraphics[width=2\columnwidth]{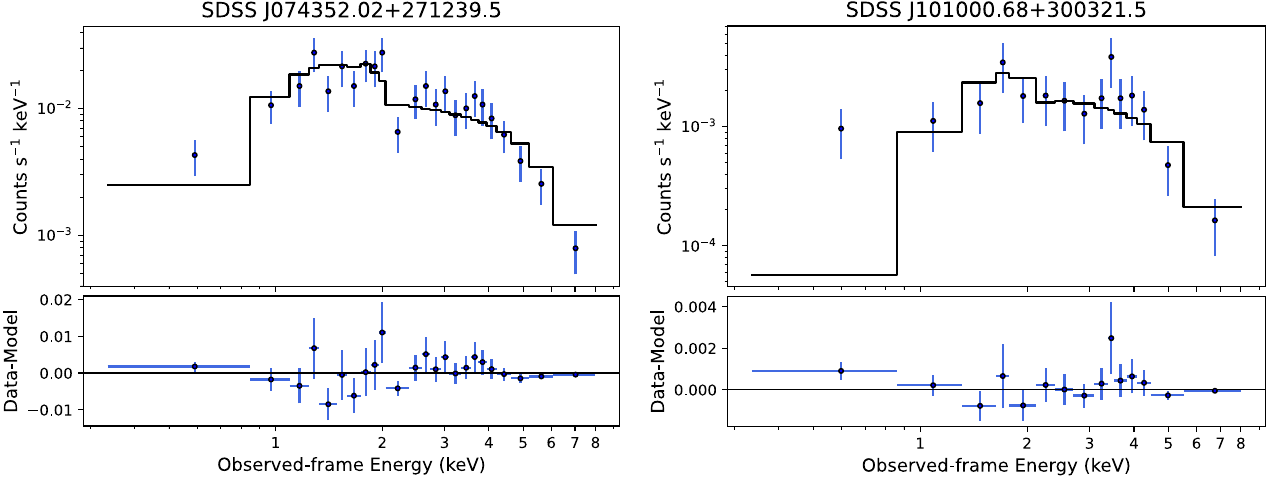}
  \caption{\label{second}Power-law fit with Galactic absorption to the observed-frame 0.35-8/(1+z) keV spectra of two Chandra SEAMBHs which shows harder spectrum above 2 keV.} 
  \label{fig:Figure2b}
\end{figure*}
\begin{figure*}
  \includegraphics[width=1.1\columnwidth]{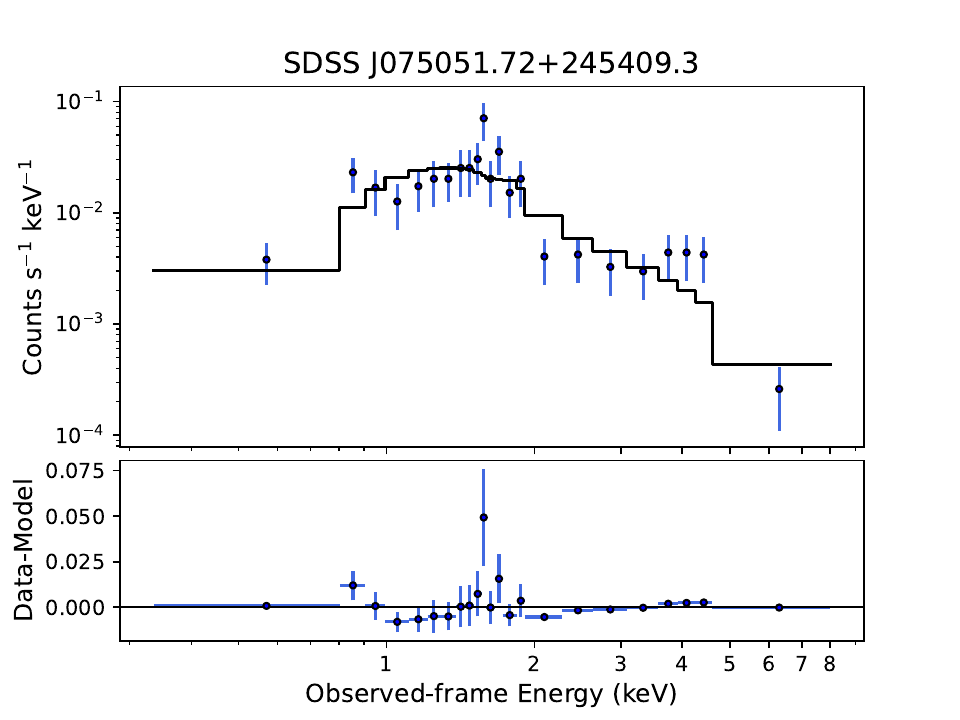}
  \caption{\label{second}Power-law fit with Galactic absorption to the observed-frame 0.35-8/(1+z) keV Chandra spectrum of one Archived SEAMBH analyzed in this paper.} 
  \label{fig:Figure2c}
\end{figure*}
\end{subfigures}

\subsection{Optical spectral analysis}\label{section3b}
We obtained the optical spectra of the eight Chandra SEAMBHs almost simultaneously with the X-ray observations, using the Lijiang and/or CAHA observatories. Multiple observations with two sets of spectra were taken within 2-8 days of the Chandra observations. We selected the best-quality spectra closest in time to the Chandra observation.

For the six objects observed at Lijiang 2.4m telescope (listed in Table \ref{tab:table1}),
their spectra were taken under the same settings of the grism and slit width
as they were observed in \cite{Hu+2015} (for IRAS 04416$+$1215), \cite{Du+2015}
(for SDSS J080131.58$+$354436.4), and \cite{Du+2018} (for the
other four objects), respectively. For the two objects observed at CAHA 2.2m
telescope, the spectra were taken using the Calar Alto Faint Object
Spectrograph with Grism G-200 and a slit set at a width of 3$\farcs$0, as
described in \cite{Hu+2021}. For all the objects, the same comparison stars
were taken simultaneously by rotating the slit, as they were monitored for 
reverberation mapping measurements. The data were reduced firstly by the
standard procedures, including bias-removal, flat-field correction, wavelength
calibration, and spectral extraction. Then the flux calibration was performed
using a sensitivity function obtained from the simultaneously observed
comparison star spectrum as in the previous reverberation mapping observations
(see, e.g., \citealt{Hu+2021} for the details of the method). Thus the flux here
can be compared directly with those in the light curves of their previous
reverberation mapping observations \citep{Hu+2015,Du+2015,Du+2018}.

We dereddened the mean spectra and fit a slope to the continuum around the H$\beta$ region. The slope was extrapolated to get the monochromatic luminosities at a rest-frame wavelength of 2500 $\angstrom$ luminosity for each target. 

\begin{landscape}
\begin{table}
\scriptsize
 \caption{X-ray and optical-UV properties of Chandra SEAMBHs and SDSSJ075051.72+245409.3 from Archived SEAMBHs.}
 \label{tab:table3}
 \begin{tabular}{lcccccccccccc}
 \hline
Object  &   $\Gamma_{\rm 0.35-8~ keV}$ &  $f_{\rm 0.35-8~ keV}$  & W-stat/DOF &  $\Gamma_{\rm 2-8~ keV}$  & $f_{\rm 2-8~ keV}$ &  W-stat/DOF & $f_{\rm 2~ keV}$   &   log $L_{\rm 2~ keV}$ &  log $L_{2500\angstrom}$ & log $\nu L_{5100\angstrom}$ & $\alpha_{\rm ox}$   & $\Delta\alpha_{\rm ox}$\\
 &  & $\rm (erg~ s^{-1} cm^{-2})$ &  &  &$\rm (erg~ s^{-1} cm^{-2})$ & & $\rm (erg~ s^{-1} cm^{-2}~ keV^{-1})$  & $\rm (erg~ s^{-1}~ Hz^{-1})$ & $\rm (erg~ s^{-1}~ Hz^{-1})$ & $\rm (erg~ s^{-1})$ & & \\
 (1) & (2) & (3) & (4) & (5) & (6) & (7) & (8) & (9) & (10) & (11) & (12) & (13) \\
 \hline
IRAS04416+1215  &   2.13$\pm0.05$    & 6.47E-12 &   41.3/38    &   2.03$\pm0.12$    &   2.67E-12    &   29.5/32    &   1.07E-12    &   25.90   &   29.67   &   44.47   &   -1.45   &   -0.02   \\
SDSSJ074352.02+271239.5$^\ddagger$  &   0.93$\pm0.13$    & 8.09E-13 &   22.0/21    &   1.98$\pm0.53$    &   8.97E-13    &   10.6/15    &   4.01E-13    &   26.41   &   30.57   &   45.37   &   -1.60   &   -0.05   \\
SDSSJ080131.58+354436.4  &   3.20$\pm0.07$    & 3.77E-12 &   53.3/30    &   2.48$\pm0.18$    &   4.55E-13    &   10.3/7    &   2.68E-13    &   25.92   &   29.17   &   43.97   &   -1.24   &   0.11    \\
SDSSJ083553.46+055317.1  &   2.69$\pm0.05$    & 3.08E-12 &   87.2/72   &   2.16$\pm0.13$    &   7.38E-13    &   20.5/28    &   3.55E-13    &   26.17   &   29.69   &   44.44   &   -1.35   &   0.08    \\
SDSSJ084533.28+474934.5  &   2.49$\pm0.10$    & 3.85E-13  &   22.8/22    &   2.36$\pm0.22$    &   1.13E-13    &   8.3/9    &   6.71E-14    &   25.80   &   29.37   &   44.53   &   -1.37   &   0.01    \\
SDSSJ093302.68+385228.0  &   2.43$\pm0.10$    & 6.84E-13 &   26.7/26    &   2.27$\pm0.22$    &   2.03E-13    &    27.9/28    &   1.04E-13    &   25.51   &   29.36   &   44.31   &   -1.48   &   -0.09   \\
SDSSJ100402.61+285535.3  &   2.53$\pm0.08$    & 1.64E-12 &   95.9/91    &   2.10$\pm0.16$    &   5.06E-13    &   13.6/12    &   2.58E-13    &   26.45   &   30.54   &   45.52   &   -1.57   &   -0.02   \\
SDSSJ101000.68+300321.5$^\ddagger$  &   0.54$\pm0.24$    & 1.16E-13 &   14.9/13    &   1.54$\pm1.03$    &   1.31E-13    &   8.4/10    &   4.26E-14    &   25.45   &   29.81   &   44.76   &   -1.67   &   -0.23   \\ \midrule
SDSSJ075051.72+245409.3  &   2.38$\pm0.19$    & 7.46E-13 &   49.8/50    &   2.39$\pm0.31$    &   2.61E-13    &   38.7/32    &   1.71E-13    &   26.45   &   30.41   &   45.33   &   -1.52   &    0.01    \\
\hline
 \end{tabular}
\\
 $^\ddagger$We fit the observed-frame 2/(1+z) to 8/(1+z) keV spectra of SDSS J074352.02+271239.5 and SDSS J101000.68+300321.5 using a power-law model incorporating both Galactic and intrinsic absorption. The best-fit model indicates intrinsic absorption, with $N_{\rm H} = \rm 3.5\times 10^{22}~ cm^{-2}$ and $\rm 3.1\times 10^{22}~ cm^{-2}$ for SDSS J074352.02+271239.5
and SDSS J101000.68+300321.5, respectively.
\end{table}

\end{landscape}

\begin{table*}
 \caption{Best-fit values for $<2\rm~ keV$ X-ray spectral fitting.}
 \label{tab:table4}
 \begin{tabular}{lclll}
 \hline
Object	&	Model & $\Gamma_{\rm 0.35-2 keV}$  & 	$f_{\rm 0.35-2~ keV}$    &   W-stat/DOF\\
    &   &   &  $\rm (erg~ s^{-1} cm^{-2})$  &    \\
 \hline
IRAS04416+1215	&	1	&  2.52$\pm$0.14   & 4.84E-12   & 30.9/36 \\
$^\dagger$SDSSJ074352.02+271239.5 & ...	&	11.98 &   2.31E-13 & 10.7/11\\
SDSSJ080131.58+354436.4 & 1	&	3.51$\pm$0.11 &   4.19E-12 & 61.9/54\\
SDSSJ083553.46+055317.1 & 1	&	3.12$\pm$0.11 &   3.18E-12 & 36.4/43\\
$^\dagger$SDSSJ084533.28+474934.5 & ...	&	2.36 &   2.56E-13 & 15.2/17\\
$^\dagger$SDSSJ093302.68+385228.0 & ...	&	2.27 &   4.46E-13 & 19.8/16\\
SDSSJ100402.61+285535.3 & 1	&	3.27$\pm$0.18 &   1.82E-12 & 25.8/26\\
SDSSJ101000.68+300321.5 & 1	&	3.10$\pm$1.10 &   4.08E-14 & 6.6/6\\ \midrule
$^\dagger$SDSSJ075051.72+245409.3 & ... &   2.39 &   4.74E-13 & 5.8/6 \\
 \hline
 \end{tabular}

 \footnotesize{Note-$^\dagger$For these targets extending 2-8 keV Galactic-absorbed power-law model (see Table \ref{tab:table3}) to $<2\rm ~ keV$ energies gives the best fitting results.}
\end{table*}

\section{Measured \& derived quantities}\label{section4}
\begin{enumerate}
\item The broadband photon index ($\Gamma_{\rm 0.35-8~ kev}$) is obtained from the best-fit model at the observed-frame 0.35-8/(1+z) keV energy band and the $\Gamma_{\rm 2-8~ keV}$ in the observed-frame 2/(1+z) to 8/(1+z) keV energy band. The absorption-corrected rest-frame fluxes ($f_{\rm 0.35-8~ keV}, f_{\rm 2-8~ keV}$) were also obtained from these best-fit models. We also obtained the monochromatic flux density at rest-frame 2 keV ($f_{\rm 2keV}$) from the fitting of $\rm 2-8~ keV$ energy band. The flux density ($F_{\nu}$) is converted to luminosity using $L_{\nu} = 4 \pi  D_{L}^2  F_{\nu}/(1+z)$ \citep[e.g.,][]{Hogg+2002}.

\item We used the following equation to derive $\alpha_{\rm ox}$
    \begin{equation}
        \alpha_{\rm ox} = \frac{{\rm log} (f_{\rm 2~keV}/f_{2500 \angstrom}) }{{\rm log} (\nu_{\rm 2~keV}/\nu_{\rm 2500 \angstrom})} = 0.3838~ {\rm log} (f_{\rm 2~keV}/f_{2500 \angstrom})
    \label{eq:eq1}
    \end{equation}
where, $f_{\rm 2~keV}$ and $f_{2500 \angstrom}$ are X-ray (2 keV) and UV (2500$\angstrom$) monochromatic flux densities, respectively. 
A highly significant anti-correlation between $\alpha_{\rm ox}$ and $L_{2500\angstrom}$ provides a method to predict $\alpha_{\rm ox}$ based on $L_{2500\angstrom}$ measurement. We define $\Delta \alpha_{\rm ox}$ as the difference between the measured $\alpha_{\rm ox}$ and predicted from \cite{Steffen2006} relation given as
\begin{equation}
    \alpha_{\rm ox} = -(0.137\pm0.008)~ {\rm log}~  L_{2500 \angstrom} + (2.638\pm0.240). 
\label{eq:eq2}
\end{equation}

\item The black hole mass ($M_{\rm BH}$) is the H$\beta$ reverberation-mapped mass. We use two accretion rate parameters: (a) The dimensionless accretion rate parameter $\dot{\mathscr{M}}$ is defined as
    \begin{equation}
    \dot{\mathscr{M}} = 20.1 \left( \ell_{44}/\cos i \right)^{3/2} m_7^{-2},
    \label{eq:eq3}
    \end{equation}
where, $m_7 = M_{\rm BH}$/$10^7 M_{\sun}$, $\ell_{44} = L_{5100\angstrom}/10^{44}$ is in units of $\rm erg~ s^{-1}$ and $i$ is inclination angle to the line of sight (we assumed a typical value of $\cos i = 0.75$ for type-1 AGNs), (b) The Eddington ratio is defined as the ratio of bolometric luminosity ( $L_{\rm Bol}$) and Eddington luminosity ($L_{\rm Edd}$),
  \begin{equation}
  L_{\rm Bol}/L_{\rm Edd} = \frac{9.26~ L_{5100\angstrom}}{1.5 \times 10^{45}~ m_7~ {\rm erg~ s^{-1}}}. 
 \label{eq:eq4}
 \end{equation}

For our Chandra SEAMBHs, we used new measurements of $\Gamma_{\rm 2-8~ keV}$, $L_{2~ \rm keV}$ and $L_{2500\angstrom}$ and calculated $\alpha_{\rm ox}$ using equation (\ref{eq:eq1}). All the X-ray and optical-UV measurements for the Archival SEAMBHs, super-Eddington, and sub-Eddington samples were adopted from \cite{Liu+2021}. The monochromatic luminosity at $5100~ \angstrom$ ($L_{5100\angstrom}$) and black hole properties like $M_{\rm BH}$ and $\dot{\mathscr{M}}$ were taken from \cite{Du+Wang2019} for all samples. \cite{Du+Wang2019} computes $\dot{\mathscr{M}}$ using $L_{5100~\angstrom}$ (see equation \ref{eq:eq3}), whereas \cite{Liu+2021} compute $\dot{\mathscr{M}}$ using $L_{2500\angstrom}$. Our choice of using the \cite{Du+Wang2019} prescription of $\dot{\mathscr{M}}$ results in two objects from \cite{Liu+2021} super-Eddington sample, PG 0953+414 \& NGC 4748, to have $\dot{\mathscr{M}}<3$. We count these two targets in the sub-Eddington sample. One target in the \cite{Liu+2021} sub-Eddington sample, PG 0026+129, now has $\dot{\mathscr{M}}>3$, and we count it in the super-Eddington sample. We calculate the Eddington ratio for all samples using equation \ref{eq:eq4}, where we make a conservative choice of using \cite{Richards+2006} bolometric correction factor of 9.26 to estimate the bolometric luminosity (see Section \ref{section6} for further discussion).

\end{enumerate}

\section{Results}\label{section5}
\begin{figure}
   \includegraphics[width=\columnwidth]{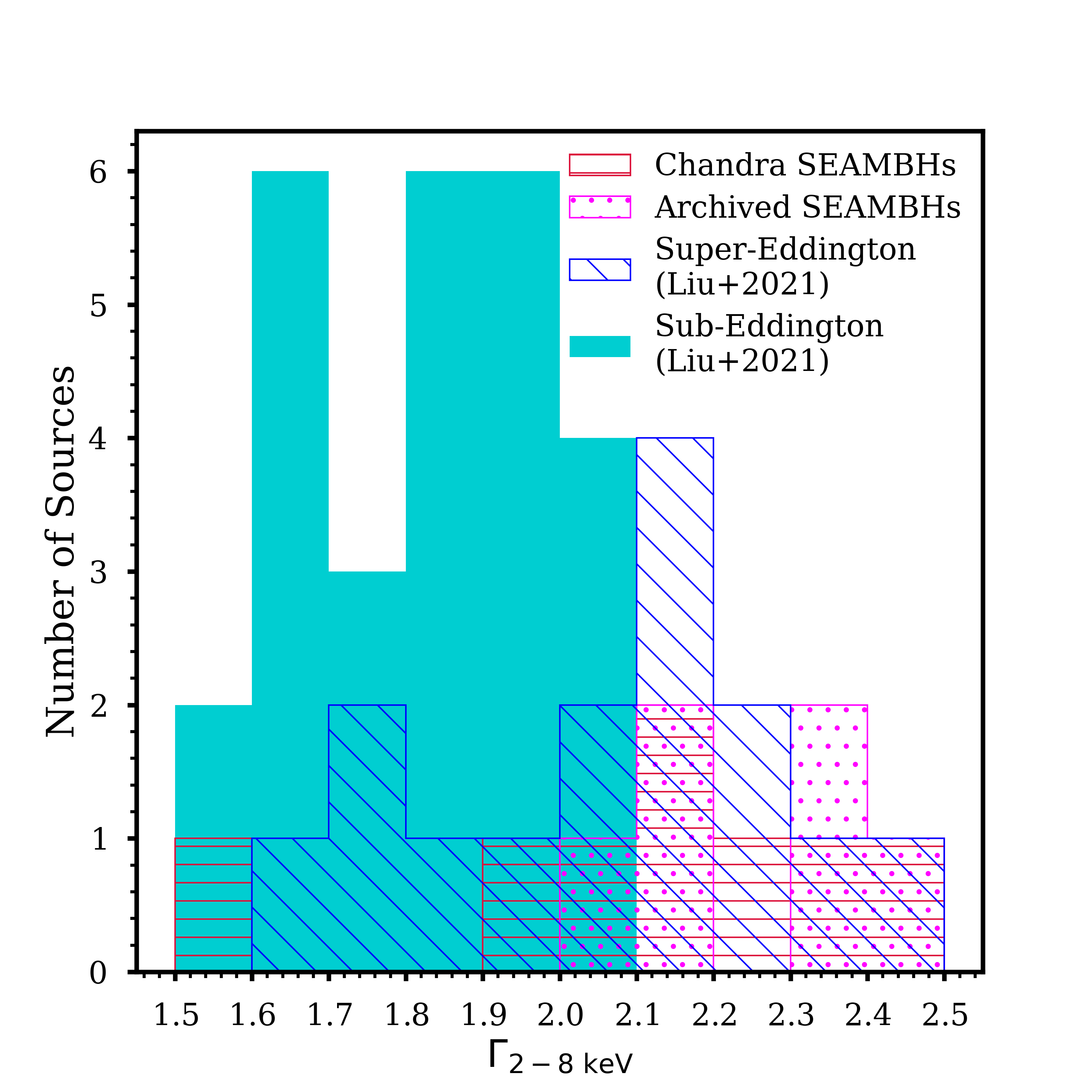}
   \caption{Histogram of the distribution of $\Gamma_{\rm 2-8~ keV}$ for all samples. The sub-Eddington sources primarily have $\Gamma_{\rm 2-8~ keV}<2$ with a mean value of 1.81, whereas, the super-Eddington and extreme SEAMBHs samples span a wider range in $\Gamma_{\rm 2-8~ keV}$ and have a mean value of 2.11.}
   \label{fig:Figure3}
\end{figure}

The distribution of 2-8 keV photon indices (Figure \ref{fig:Figure3}) shows that extreme SEAMBHs typically exhibit steep photon indices, with $\Gamma_{\rm 2-8~ keV}>2$. However, two Chandra SEAMBHs, SDSS J074352.02+271239.5 \& SDSS J101000.68+300321.5, have $\Gamma_{\rm 2-8~ keV}<2$. SDSS J101000.68+300321.5 has a low photon count (broadband photon count of 72) and, therefore, the spectral fit is likely inaccurate. The $\Gamma_{\rm 2-8~ keV}$ for the extreme SEAMBHs (${\rm log}\dot{\mathscr{M}} > 1.5$) ranges between 1.54 and 2.48 and has a mean value of 2.17 (2.22 excluding SDSS J101000.68+300321.5). In contrast, the mean $\Gamma_{\rm 2-8~ keV}$ the for 15 super-Eddington AGNs ($0.47 < {\rm log}\dot{\mathscr{M}} < 1.5$) is 2.06, which is clearly distinct from the mean $\Gamma_{\rm 2-8~ keV} $ of 1.81 for the sub-Eddington AGNs.   

\begin{figure}
   \includegraphics[width=\columnwidth]{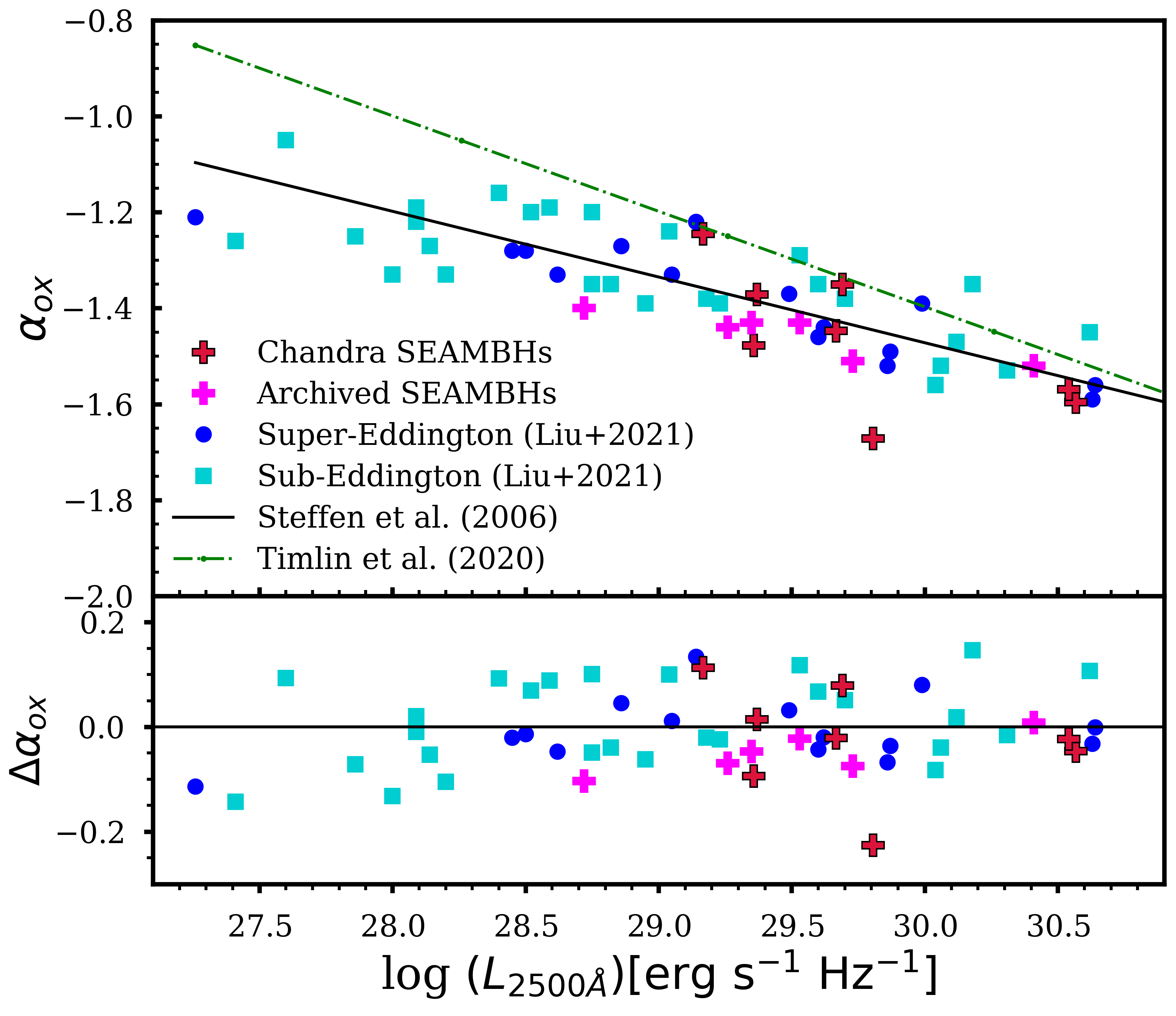}
   \caption{The top panel shows that $\alpha_{\rm ox}$ becomes more negative with increasing $L_{\rm 2500 \angstrom}$ following the Steffen et al. (2006) relation represented by the black solid line. Our sample shows a large deviation from Timlin et al. (2020) relation, represented by the green dash-dot line, especially at the low luminosity end. The bottom panel plots $\Delta \alpha_{\rm ox}$ as a function of $L_{\rm 2500 \angstrom}$. A $\Delta \alpha_{\rm ox}$ less than $-$0.2 implies significant X-ray weakness.}
   \label{fig:Figure4}
\end{figure}

Figure \ref{fig:Figure4} illustrates the $\alpha_{\rm ox}$ vs $L_{2500\angstrom}$ plane. The $\alpha_{\rm ox}$ values of extreme SEAMBHs range between -1.67 and -1.24, following the best-fit linear regression from \cite{Steffen2006} as expressed in equation (\ref{eq:eq2}). The mean $\alpha_{\rm ox}$ values are -1.46, -1.38 and -1.32 for the extreme SEAMBHs, super-Eddington and sub-Eddington samples, respectively. $\alpha_{\rm ox}$ tends to steepen with increasing $L_{2500\angstrom}$, consistent with previous studies. We also plot the $\alpha_{\rm ox}-L_{2500\angstrom}$ best-fit relationship from \cite{Timlin+2020}, given by $\alpha_{\rm ox} = (-0.199\pm0.011) {\rm log_{10}} (L_{2500}) + (4.573\pm 0.333)$. Notably, this slope is significantly steeper than that of \cite{Steffen2006}.  \cite{Timlin+2020} uses a Bayesian linear regression method developed by \citep{Kelly2007}, while \cite{Steffen2006} utilizes a bivariate data-analysis method presented in \citep{Isobe+1990} for their best-fit relations (see their respective papers for further details). Figure \ref{fig:Figure4} shows that the \cite{Timlin+2020} relationship does not fit our sample very well, particularly at the low luminosity end. \cite{Timlin+2020} derived their relation from a sample of 753 quasars with $29.3\lesssim{\rm log}~ L_{2500\angstrom}\lesssim31.6$, spanning only two orders of magnitude. In contrast, the \cite{Steffen2006} relation is based on a sample of 293 quasars with $28\lesssim{\rm log}~ L_{2500\angstrom}\lesssim33$, covering four orders of magnitude. Given that the luminosity of our sample spans three orders of magnitude, $27.2<{\rm log}~ L_{2500\angstrom}<30.7$, it is more appropriate to use the \cite{Steffen2006} relationship to calculate $\Delta \alpha_{\rm ox}$.

\begin{figure}
   \includegraphics[width=\columnwidth]{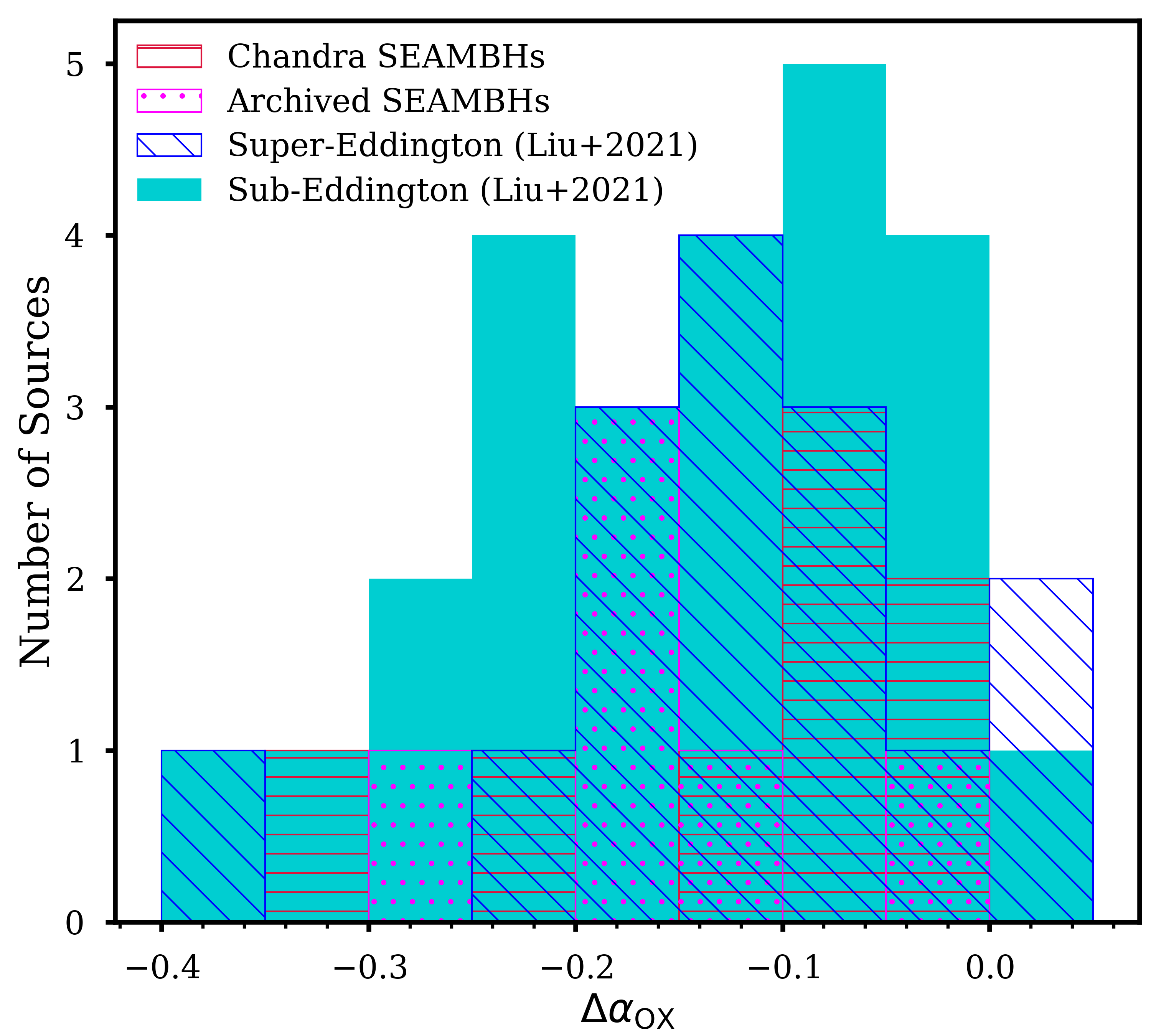}
   \caption{Histogram of the distribution of $\Delta \alpha_{\rm ox}$. Except for one extreme SEAMBH, SDSS J101000.68+300321.5, all sources appear X-ray normal.}
   \label{fig:Figure5}
\end{figure}

The bottom panel of Figure \ref{fig:Figure4} shows the residual between the measured $\alpha_{\rm ox}$ and the expected value from \cite{Steffen2006} relation (as given in equation \ref{eq:eq2}), plotted against $L_{2500\angstrom}$. We plot the distribution of $\Delta \alpha_{\rm ox}$ in Figure \ref{fig:Figure5}. \cite{Luo+2015} defines $\Delta \alpha_{\rm ox} = -0.2$ as a threshold value, below which the AGN is considered X-ray weak. Apart from one Chandra SEAMBH, $\Delta \alpha_{\rm ox}$ varies between $-$0.15 and 0.15, typical for normal X-ray emission. Only SDSS J101000.68+300321.5 appears as X-ray weak, with $\Delta \alpha_{\rm ox} = -0.23$. The limited photon count for this object hinders fitting more complex models to correct for absorption.

\begin{figure}
   \includegraphics[width=1.1\columnwidth]{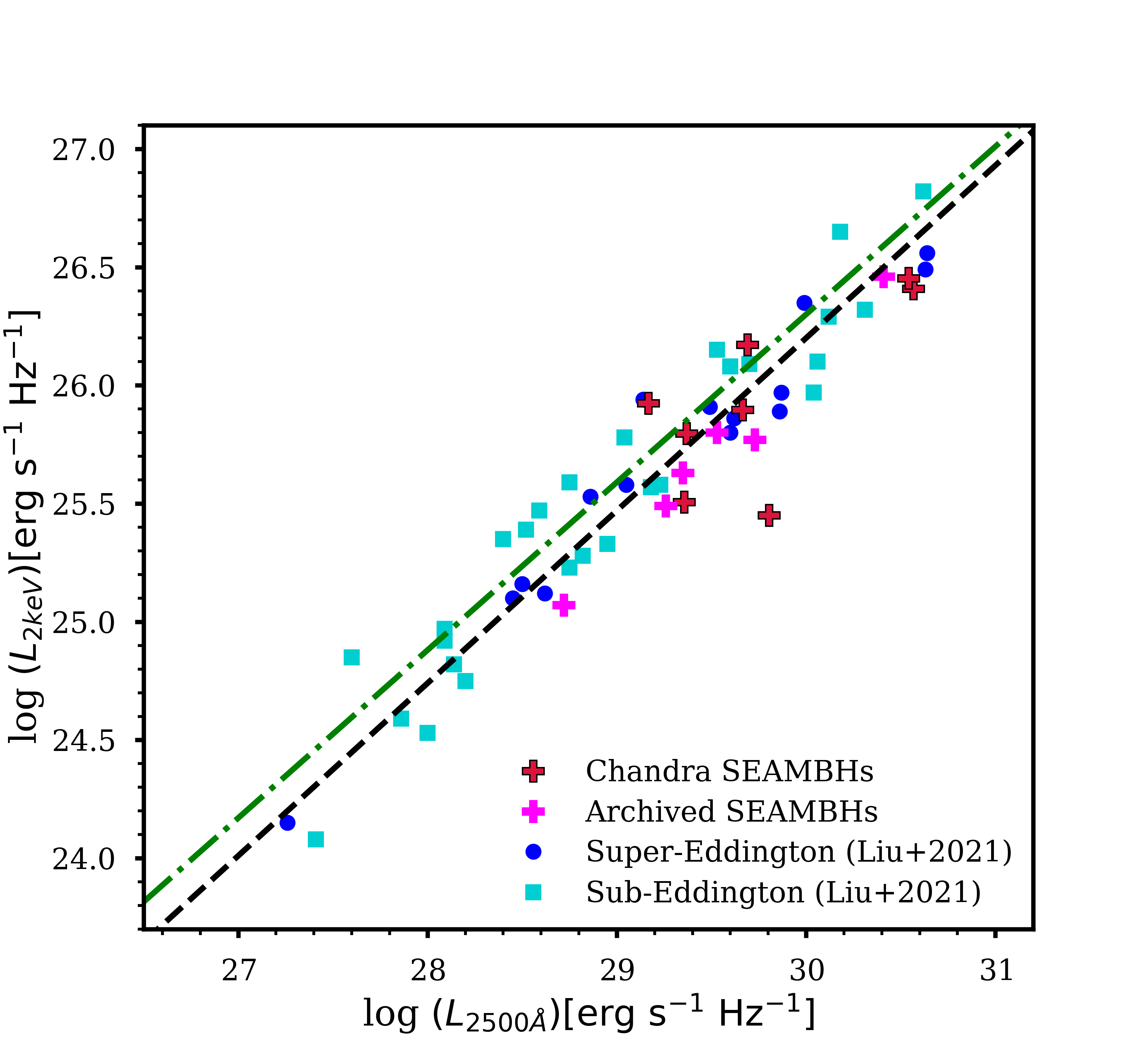}
   \caption{$L_{\rm 2 keV}$ vs $L_{\rm 2500 \angstrom}$. The black dashed (green dashed-dotted)
   line represents the best-fit relation for super-Eddington (sub-Eddington) sources from Liu et al. (2021). The plot shows that the non-linear correlation between X-ray and optical-UV luminosities is also present in extreme SEAMBHs.}
   \label{fig:Figure6}
\end{figure}

Figure \ref{fig:Figure6} plots the best-fit relation between $L_{2~\rm keV}$ and $L_{2500\angstrom}$ for the super-Eddington sample (${\rm log~} L_{2~\rm keV} = (0.73\pm0.05)~ {\rm log~} L_{2500\angstrom} + (4.3\pm1.4)$) and the sub-Eddington sample (${\rm log~} L_{2~\rm keV} = (0.71\pm0.05)~ {\rm log~} L_{2500\angstrom} + (5.0\pm1.5)$) from \cite{Liu+2021}. These best-fit relationships were derived using \texttt{LINMIX\_ERR} method \citep{Kelly2007}, which is a linear regression method that utilizes Bayesian priors for errors (refer to \citealt{Liu+2021} for further details). The slope of \citealt{Liu+2021} best-fit relation is within the error range of the typical value of 0.6$\pm$0.1 seen in previous studies \citep[e.g.,][]{Steffen2006, Lusso+2010}. 
The trend of suppression of X-ray emission in comparison to optical-UV emission holds for extreme SEAMBHs.

\subsection{$\Gamma_{\rm 2-8~ keV}$ vs accretion rate}
The $\rm 2-8~ keV$ photon index is known to soften or steepen with an increasing accretion rate. \cite{Shemmer+2008} finds a significant correlation between $\Gamma_{\rm 2-8~ keV}$ and $L_{\rm Bol}/L_{\rm Edd}$ and derived a linear relation of the form $\Gamma_{\rm 2-8~ keV} = a~ {\rm log}~ L_{\rm Bol}/L_{\rm Edd} + b$, where the slope $a = 0.31$ and intercept $b = 2.11$. Subsequently, the $\Gamma_{\rm 2-8~ keV}$-$L_{\rm Bol}/L_{\rm Edd}$ correlation has been tested across AGNs spanning a wide range of luminosities and redshifts. For instance, \cite{Risaliti+2009} finds $a=0.31,~ b=1.97$ for their full sample, \cite{Brightman+2013} reports $a=0.32,~ b =2.27$, and \cite{Fanali+2013} finds $a=0.25,~ b=2.48$. 

\begin{figure}
   \includegraphics[width=\columnwidth]{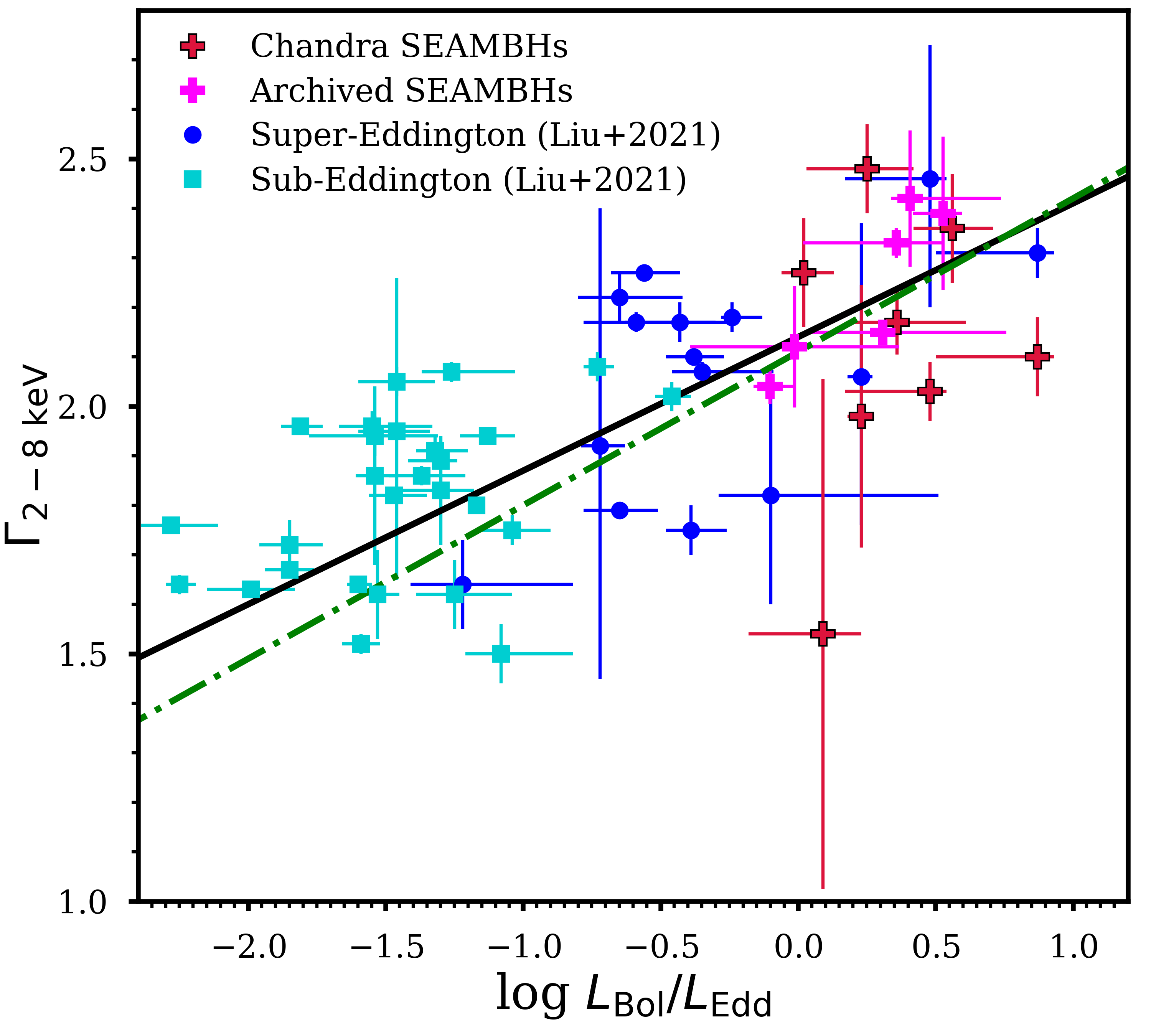}
   \caption{The plot $\Gamma_{\rm 2-8~ keV}$ vs $L_{\rm Bol}/L_{\rm Edd}$ shows that $\Gamma_{\rm 2-8~ keV}$ steepens with increasing accretion rate (Spearman r = 0.70, p = 2.42E-9). The green dashed-dotted line represents the best-fit relation from Shemmer et al. (2008) and the black solid line represents the best-fit relation from Liu et al. (2021).}
   \label{fig:Figure7}
\end{figure}

Figure \ref{fig:Figure7} illustrates the relationship between $\Gamma_{\rm 2-8~ keV}$ and $L_{\rm Bol}/L_{\rm Edd}$.
It plots the best-fit relation for the full sample from \cite{Liu+2021} as $\Gamma_{\rm 2-8~ keV} = (0.27\pm0.04)~ {\rm log}~ L_{\rm Bol}/L_{\rm Edd} + (2.14\pm0.04)$ in black, alongside the the \cite{Shemmer+2008} relation, $\Gamma_{\rm 2-8~ keV} = (0.31\pm0.01)~ {\rm log}~ L_{\rm Bol}/L_{\rm Edd} + (2.11\pm0.01)$, represented by a green dot-dash line. Both best-fit relationships were derived using linear regression techniques that account for measurement errors. \cite{Liu+2021} employed the \texttt{LINMIX\_ERR} method \citep{Kelly2007}, a Bayesian approach incorporating a prior distribution for errors. In contrast, \cite{Shemmer+2008} utilized the bivariate correlated errors and scatter method \citep{AkritasBershady1996}, which assumes random errors with no correlation. The extreme SEAMBHs follow the trend of steeping $\Gamma_{\rm 2-8~ keV}$ as the Eddington ratio increases. The full sample shows a strong correlation between $\Gamma_{\rm 2-8~ keV}$ and $L_{\rm Bol}/L_{\rm Edd}$, with a Pearson $r$-coefficient of 0.70 and a probability $p = 2.37E-9$. Removing the sub-Eddington sample weakens the Pearson correlation to $r=0.49, p = 0.01$. We also tested the Spearman non-parametric correlation between the $\Gamma_{\rm 2-8~ keV}$ and the Eddington ratio, yielding a $r$-coefficient = 0.70 and a probability $p = 2.42E-9$ for the full sample. Without the sub-Eddington sample, the Spearman correlation decreases to $r=0.48, p = 0.01$. A weak correlation between $\Gamma_{\rm 2-8~ keV}$ and $L_{\rm Bol}/L_{\rm Edd}$ is observed for the 14 extreme SEAMBHs, with Pearson (Spearman) $r=0.29 (0.33), p=0.31 (0.25)$. This may be be attributed to the shortcoming of a constant bolometric correction factor of 9.26 for estimating the bolometric luminosities in these highly accreting systems or, the small sample size, or the narrow range of $L_{\rm Bol}/L_{\rm Edd}$ and $\Gamma_{\rm 2-8~ keV}$ spanned by the sample.

\begin{figure}
   \includegraphics[width=\columnwidth]{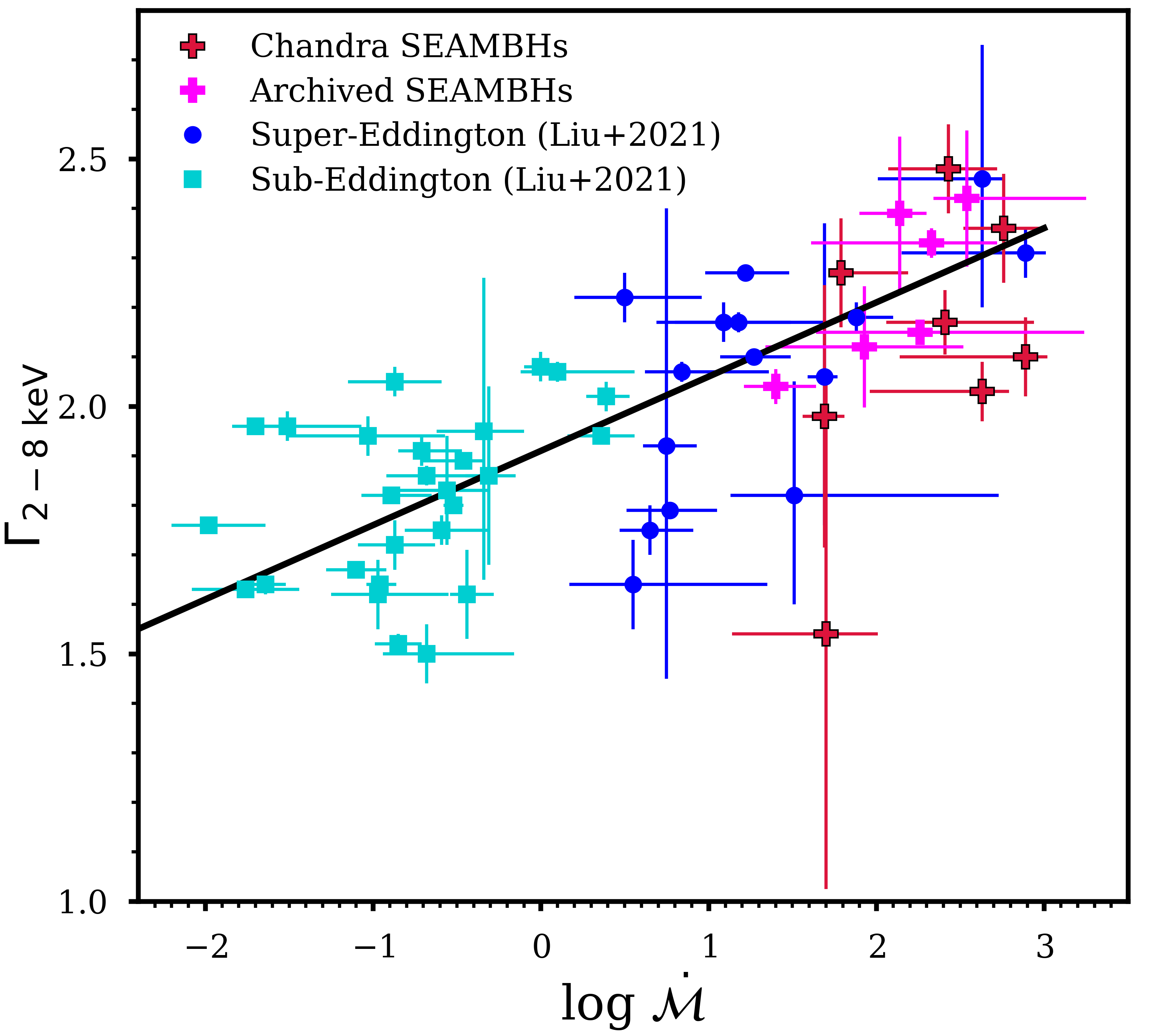}
   \caption{$\Gamma_{\rm 2-8~ keV}$ vs $\dot{\mathscr{M}}$. The $\rm 2-8~ keV$ photon index shows strong correlation  with $\dot{\mathscr{M}}$ (Spearman r = 0.72, p = 4.92E-10). The black solid line represents the best-fit relation from Liu et al. (2021).}
   \label{fig:Figure8}
\end{figure}

Using $\dot{\mathscr{M}}$ as an accretion rate indicator is complementary to using the Eddington ratio, although these parameters are tightly correlated for AGNs. Figure \ref{fig:Figure8} plots $\Gamma_{\rm 2-8~ keV}$ vs $\dot{\mathscr{M}}$ for our full sample and displays the best-fit relation from \cite{Liu+2021} for their full sample, $\Gamma_{\rm 2-8~ keV} = (0.15\pm0.02)~ {\rm log}~ \dot{\mathscr{M}} + (1.91\pm0.03)$, in black. Again, \cite{Liu+2021} employed the Bayesian linear regression method \texttt{LINMIX\_ERR} \citep{Kelly2007}, to derive the best-fit relationship. The full sample shows a strong Pearson (Spearman) correlation between $\Gamma_{\rm 2-8~ keV}$ and $\dot{\mathscr{M}}$, with $r = 0.71$ (0.72) and $p = 5.80E-10$ ($4.92E-10$). This correlation remains strong when the sub-Eddington sample is excluded, with Pearson (Spearman) $r=0.57 (0.55) and p = 1.3E-03 (1.9E-03)$. $\Gamma_{\rm 2-8~ keV}$ and $\dot{\mathscr{M}}$ correlate strongly with Pearson (Spearman) $r=0.46$ (0.41) and $p = 0.10$ (0.14) for the extreme SEAMBHs. Our results confirm that the correlation between $\Gamma_{\rm 2-8~ keV}$ and $\dot{\mathscr{M}}$ extends to extreme SEAMBHs.

\subsection{$\alpha_{\rm ox}$ vs accretion rate and black hole mass}
\begin{figure}
   \includegraphics[width=\columnwidth]{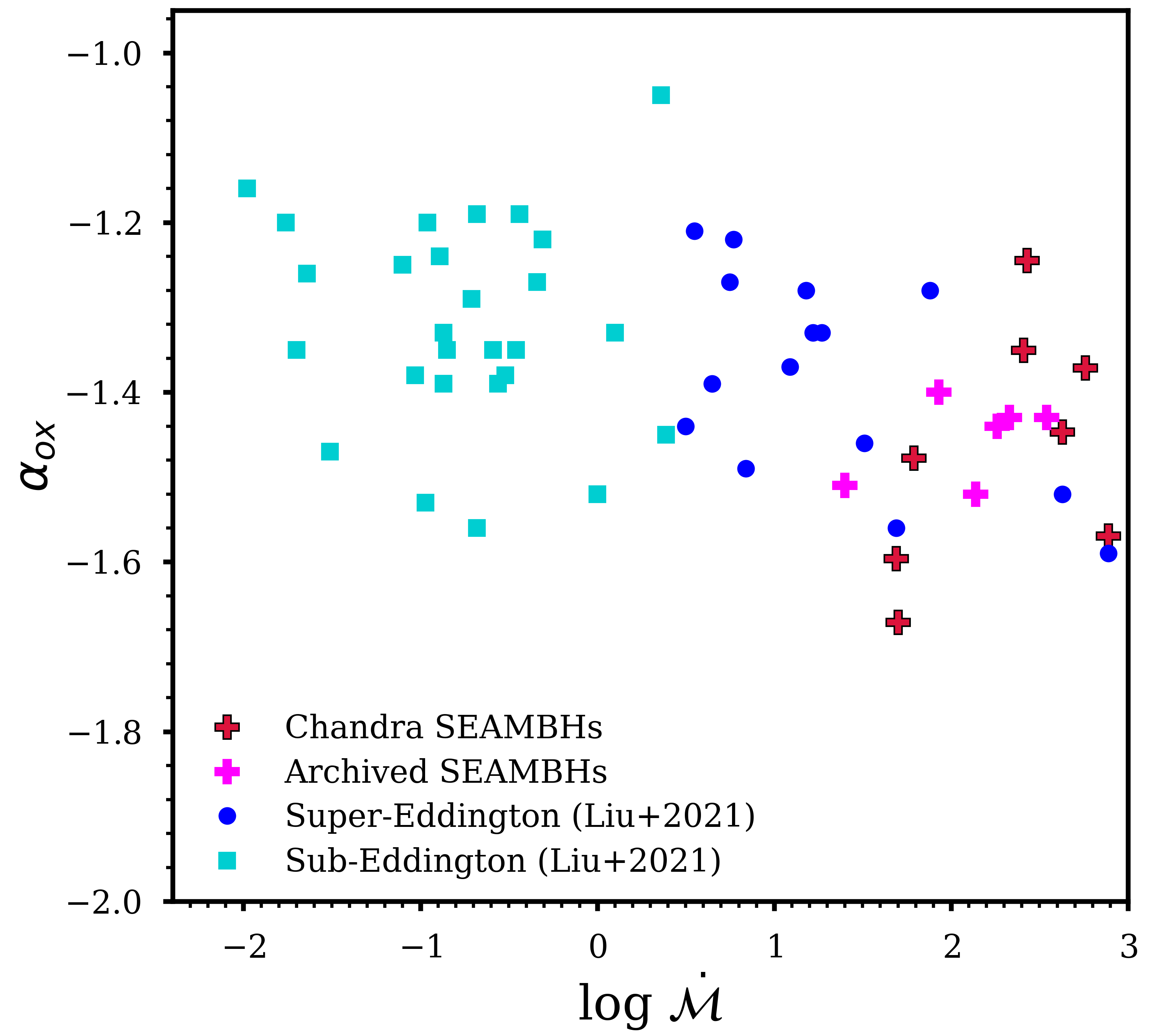}
   \caption{$\alpha_{\rm ox}$ vs $\dot{\mathscr{M}}$. The plot shows a weak anti-correlation with Spearman r = -0.45, p = 5.1E-4 and weakens further if the sub-Eddington sample is excluded. }
   \label{fig:Figure9}
\end{figure}

The physics behind the disk-corona connection responsible for the observed steepening of $\alpha_{\rm ox}$ with increasing $L_{2500\angstrom}$ is not well understood (see Section \ref{section6} for further discussion). Past studies have investigated the connection between $\alpha_{\rm ox}$ and black hole mass and accretion rate. A significant correlation is evident between $\alpha_{\rm ox}$ and $M_{\rm BH}$ \citep{Done+2012,Chiaraluce+2018}. However, there is only a weak or non-existent correlation between $\alpha_{\rm ox}$ and $L_{\rm Bol}/L_{\rm Edd}$ \citep{Vasudevan+Fabian2007,Shemmer+2008,Fanali+2013}, although \cite{Grupe+2010} claims a strong correlation between them. The dependence of $L_{\rm Bol}$ on $L_{2500\angstrom}$ might be responsible for a weak correlation between $\alpha_{\rm ox}$ and $L_{\rm Bol}/L_{\rm Edd}$ \citep{Shemmer+2008}. \cite{Fanali+2013} finds a significant anti-correlation between $\alpha_{\rm ox}$ and $\dot{\mathscr{M}}$ for a sample of 71 type-1 AGNs using canonical single-epoch black hole masses to estimate $\dot{\mathscr{M}}$, which we now know suffer from accretion rate effects. \cite{Castello+2017} claims an anti-correlation between $\alpha_{\rm ox}$ and $\dot{\mathscr{M}}$ for 31 high-accretion rate sources ($\dot{\mathscr{M}}$>10) in their sample of 59 AGNs, using RM masses and an updated R-L relationship \citep{Du+2016} for high-accretion rate targets, although the correlation is limited by small number statistics. We examine the correlation between $\alpha_{\rm ox}$ and $\dot{\mathscr{M}}$ for our sample. Figure \ref{fig:Figure9} shows a slight anti-correlation (Pearson $r=-0.45, p = 4.9E-4$) for the full sample which weakens to Pearson (Spearman) $r= -0.384 (-0.381), p=3.99E-2 (4.16E-2)$ when only $\dot{\mathscr{M}}$>3 AGNs are considered. 

\begin{figure}
   \includegraphics[width=\columnwidth]{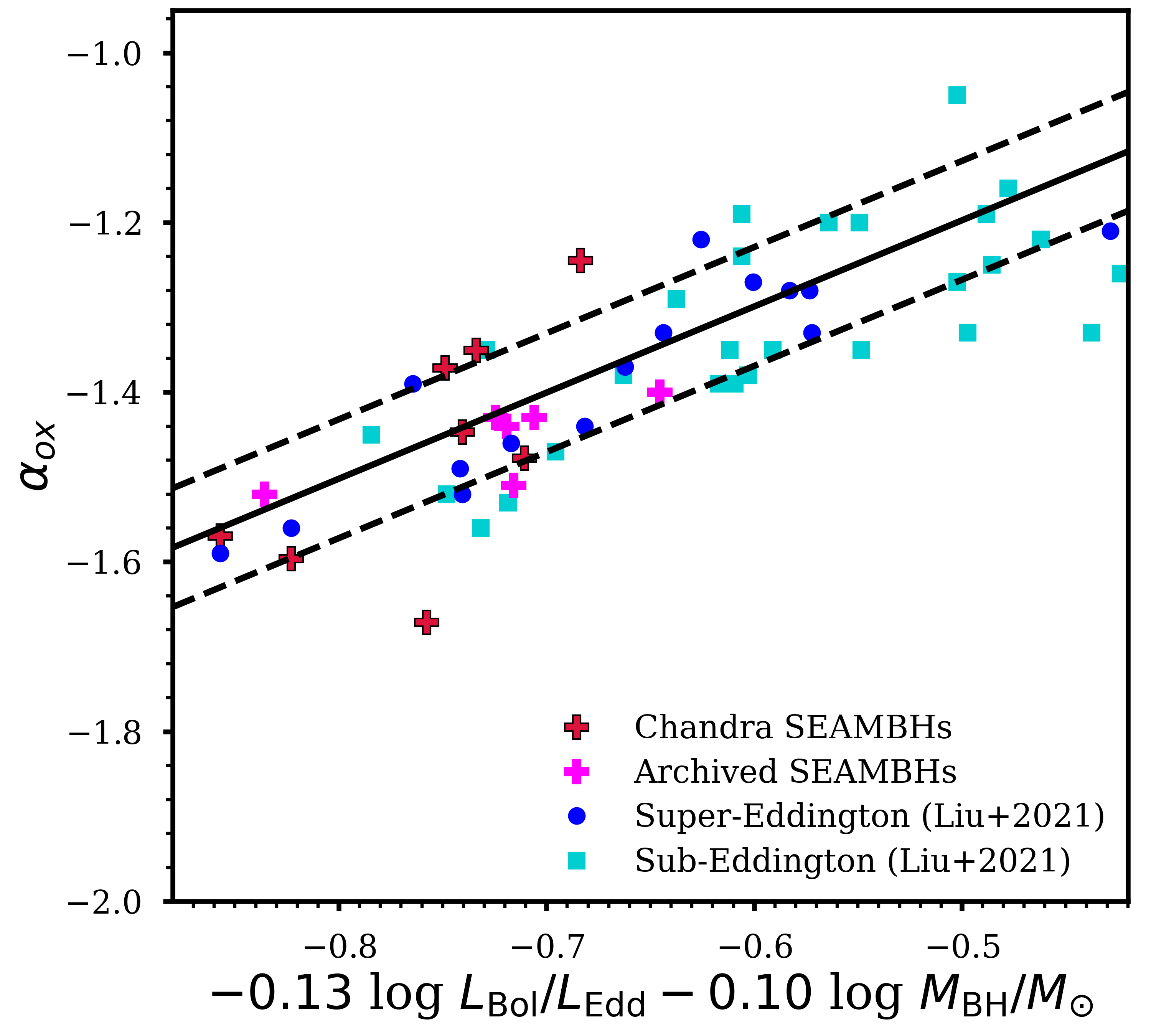}
   \caption{$\alpha_{\rm ox}$ as a function of both $\dot{\mathscr{M}}$ and $M_{\rm BH}$. The solid black line represents the best-fit relation from Liu et al. (2021) and the dashed line represents the 0.07 scatter around it.}
   \label{fig:Figure10}
\end{figure}

More recently, \cite{Liu+2021} argue that $\alpha_{\rm ox}$ depends on both $L_{\rm Bol}/L_{\rm Edd}$ and $M_{\rm BH}$, establishing a non-linear relationship of the form $\alpha_{\rm ox} = (-0.13\pm0.01)~ {\rm log}~ L_{\rm Bol}/L_{\rm Edd} - (0.10\pm 0.01)~ {\rm log}~ M_{\rm BH}- (0.69\pm0.09)$. They employed the Python package \texttt{emcee} \citep{ForemanMackey+2013}, a tool for multivariate linear regression with Bayesian inference, to obtain this best-fit relationship. More investigation is needed to decipher if this relation is fundamental or just a secondary manifestation of the $\alpha_{\rm ox}$-$L_{2500\angstrom}$ relationship. Figure \ref{fig:Figure10}  presents an edge-on view of this relationship and shows that extreme SEAMBHs fall within the 0.07 scatter of the $\alpha_{\rm ox}-L_{\rm Bol}/L_{\rm Edd}-M_{\rm BH}$ relation, with the exception of one X-ray weak Chandra SEAMBH. 

\section{Discussion \& Conclusions}\label{section6}

One version of the disk-corona model assumes a thin \cite{Shakura+Sunyaev1973} disk tightly coupled with a plane-parallel corona. The magnetic field in the accretion disk is produced by dynamo action, and the buoyancy of the magnetic fields generates magnetic loops that emerge into the corona. The loops reconnect with other loops, transferring the magnetic energy of the disk to thermal energy, thereby heating the corona. A stable corona is formed when the density of the corona reaches a certain value, allowing equilibrium between the heating by magnetic flux loops and cooling by Compton scattering.  
The coupling between the optical-UV emission from the accretion disk and the hard X-ray emission from the corona is often explained via such magnetic reconnection-heated  model \citep{Liu+2003,Cao2009}. A fraction of gravitational energy is transferred from the accretion disk to the corona through magnetic fields that inhibit the fast cooling of the corona \citep{Merloni+Fabian2001}. Magnetohydrodynamic simulations predict a decrease in the fraction of energy dissipated from the accretion disk as the disk transitions to radiation-pressure dominated in case of higher accretion rate (higher $L_{\rm Bol}/L_{\rm Edd}$)\citep[][]{Jiang+2014,Jiang+2019}. As a result, the corona becomes relatively more compact and weaker with an increasing Eddington ratio, resulting in a steeper $\alpha_{\rm ox}$. An increase in $L_{\rm Bol}/L_{\rm Edd}$ implies more soft photons from the accretion disk to cool the corona by Compton scattering, resulting in a steeper/softer $\Gamma_{\rm 2-8~ keV}$. In SEAMBHs, the inner accretion disk is perhaps geometrically thick, according to the slim disk model \citep{Abramowicz+1988,Laor+Netzer1989,Wang+1999, Wang+2013}, leading to differences in the accretion disk-corona connection. However, our analysis shows that the $\Gamma_{\rm 2-8~ keV}$-$L_{\rm Bol}/L_{\rm Edd}$($\dot{\mathscr{M}}$) and  $\alpha_{\rm ox}$-$L_{\rm 2500 \angstrom}$ (also $L_{\rm 2~ keV}$-$L_{\rm 2500 \angstrom}$) shows no dichotomy between the sub and super-Eddington sources. This could mean that the transition from geometrically thin to a slim disk is not abrupt, and the disk-corona connection remains intact in super-Eddington AGNs. Another possibility is that there is no structural difference in the accretion disk of sub- and super-Eddington sources. 

The correlations between X-ray properties and Eddington ratios studied in this paper are influenced by the choice of the bolometric correction used to estimate the bolometric luminosity. We used a correction factor of 9.26 to estimate the bolometric luminosity from $L_{5100 \angstrom}$, consistent with literature values ranging between 7 and 13 \citep[e.g.,][and references within]{Elvis+1994,Richards+2006, Runnoe+2012}. However, \cite{Jin+2012} find the bolometric correction factor of 15 for their full sample and 20 for the sample of 12 narrow-line Seyfert 1 galaxies, using a color temperature-corrected SED fitting model. Additionally, there are theoretical bolometric corrections based on the thin disk model that yield substantial differences, but they are inconsistent with empirically determined bolometric luminosies \citep[e.g.,][]{Kubota_Done2019}.

We compared our bolometric luminosity measurements with those obtained through SED fitting by \cite{Liu+2021} and the bolometric correction factor described in \cite{Netzer2019}. \cite{Liu+2021} derived bolometric luminosities by integrating the infrared-to-X-ray SED (see their Section 2.7 for details). For the three AGNs in our Chandra sample, i.e.,  IRAS04416+1215, SDSSJ074352.02+271239.5, SDSSJ100402.61+285535.3, the differences in the logarithm of bolometric luminosity measured by \cite{Liu+2021} compared to our method are 0.083, -0.017, -0.147, respectively. Notably, we found a higher level of agreement for the super-Eddington sample, with a median difference of 0.183, ranging between -0.197 and 0.413. In contrast, the median difference for the sub-Eddington sample was 0.25, ranging from -0.187 to 0.683.  It is important to emphasize that the determination of bolometric luminosities through multi-wavelength SEDs may have considerable uncertainties, especially for super-Eddington accreting quasars. This is due to various factors such as host-galaxy contamination and variability effects due to non-simultaneous SED data. Furthermore, super-Eddington accreting quasars may have significantly enhanced EUV emission compared to typical quasars, but there is no clear observational constraint on this due to the lack of data \citep{Jin+2012, Castello+2016, Kubota_Done2019}. Next, we test our choice of bolometric correction against the correction factor given in \cite{Netzer2019}. Instead of a constant of 9.26, the \cite{Netzer2019} bolometric correction factor is a function of monochromatic luminosity at $5100\angstrom$ and is expressed as $40~ [L_{\rm 5100\angstrom} (\rm observed)/10^{42} \rm erg ~ s^{-1}]^{-0.2}$. The two sets of estimates give very similar values. The agreement is slightly better for the sub-Eddington sample, with the median value of logarithmic difference being -0.0016, while it is -0.0576 for the super-Eddington sample.

Our choice of bolometric correction is conservative and does not make any enhancement in the bolometric luminosity, hence Eddington ratio, preferentially for the SEAMBHs even though that may be the case \citep{Jin+2012}. 

One of our targets, SDSS J101000.68+300321.5, exhibits a flatter/harder $\Gamma_{\rm 2-8~ keV}$ and appears as X-ray weak ($\Delta\alpha_{\rm ox}<-0.2$). It could be that the X-ray data for this object is not good enough, as indicated by large error bars, or the possibility that the X-ray emission is shielded by a puffed-up slim accretion disk. \cite{Luo+2015} explains X-ray weakness in highly accreting AGNs through an orientation effect in a slim accretion disk. When viewed at larger inclination angles, X-rays emitted from the central region may be absorbed by the puffed-up inner accretion disk \citep{Luo+2015,Ni+2018}. \cite{Liu+2019} estimate that ~15-24\% of super-Eddington AGNs should exhibit extreme X-ray variability. Our comparison sample from \cite{Liu+2021} selectively chooses the high X-ray state for the sub- and super-Eddington samples with multiple X-ray observations, thereby excluding any X-ray weak object from the sample. One recent X-ray spectroscopic study of highly accreting AGNs by \cite{Laurenti+2022} reports 29\% of the sample as X-ray weak. A dramatic change in X-ray flux is seen in some weak emission-line quasars with high accretion rates whose X-ray variability can be explained due to changes in the thickness of the accretion disk \citep{Liu+2019,Ni+2020}. Our target SDSS J101000.68+300321.5 could be an X-ray-weak weak-emission line quasar that has weak high-ionization lines like C{ \sc iv} $\rm \lambda 1549$. The lack of UV spectra for this target inhibits us from testing this hypothesis. Alternatively, the X-ray weakness may be caused by absorption from outflows, which  also manifest UV absorption troughs in some AGNs \citep[e.g.,][and references therein]{Kaastra+2014}.

It is worth noting that the extreme SEAMBHs studied in this paper are primarily narrow-line Seyfert 1 galaxies (NLS1s). NLS1s are believed to be accreting material at rates approaching the Eddington limit, as supported by various studies (Komossa et al. 2006; Komossa 2018; Gallo 2018; Foschini 2020). The optical spectra of NLS1s are characterized by several distinctive features, such as relatively narrow H$\beta$ emission lines, strong Fe {\sc ii} lines and weak [O {\sc iii}] lines \citep{Boroson2002}. These characteristics align with the Eigenvector 1 trends, known to correlate with accretion rate \citep{BG1992, Marziani+2001, Boroson2002, Yuan+2003, Shen_Ho2014, Sun_Shen2015}. In X-rays, NLS1s exhibit steep 2–10 keV spectra and a pronounced soft X-ray excess \citep[e.g.][]{Boller+1996, Brandt+1997,VeronCetty+2001}. The question arises: is the steepness of the 2-8 keV X-ray spectrum solely due to the extreme accretion rate, or can it be attributed to factors like coronal geometry or distinct cooling mechanisms within the corona specific to NLS1s? This intriguing question presents a potential avenue for future investigations using broad-band X-ray spectra. Recent research has shown a certain subjectivity in these classifications.  For instance, \cite{Jin+2023} indicated a connection between NLS1s with extreme accretion rates and weak-line quasars (WLQs). Additionally, \cite{Ha+2023} demonstrated that WLQs follow the same trend of C {\sc iv} versus Eddington ratio as other type-1 quasars.

In conclusion, we present new Chandra X-ray data of nine SEAMBHs. Our core sample consists of 14 SEAMBHs that have extremely high accretion rates (${\rm log}~ \dot{\mathscr{M}}>1.5$) and show the largest offset between the radius of BLR from the RM measurement and the one estimated from the canonical R-L relationship. We investigated the X-ray and optical-UV properties of these 14 extreme SEAMBHs and compared them to the sub- and super-Eddington quasars from \cite{Liu+2021}.  To mitigate errors due to variability, we took almost simultaneous X-ray and optical-UV observations for the Chandra SEAMBHs. However, it should be noted that there is an intrinsic delay between the X-ray and the optical-UV variability due to the spatial difference in the regions emitting these radiations. We used \cite{Du+Wang2019} for the black hole properties. Our results indicate that extreme SEAMBHs indeed have a steep 2-8 keV X-ray photon index and demonstrate a steeper power-law slope. They are consistent with correlation between $\Gamma_{\rm 2-8~ keV}$ and $L_{\rm Bol}/L_{\rm Edd}$ (also $\dot{\mathscr{M}}$) seen in sub- and super-Eddington accreting sources. We show that the  $\alpha_{\rm ox}$-$L_{\rm 2500 \angstrom}$ (also $L_{\rm 2~ keV}$-$L_{\rm 2500 \angstrom}$) correlation extends to the extreme SEAMBHs. The $\alpha_{\rm ox}$-$\dot{\mathscr{M}}$ correlation remains weak after the inclusion of eight extreme SEAMBHs; however, the bivariate relationship established by \cite{Liu+2021} between $\alpha_{\rm ox}$, $L_{\rm Bol}/L_{\rm Edd}$ and $M_{\rm BH}$ holds for the extreme SEAMBHs.

\section*{Acknowledgements}

We express our gratitude to the anonymous referee for providing constructive comments that enhanced the quality of the manuscript. We acknowledge support through Chandra Award Number GO9-20107X. B.L. acknowledges financial support from the National Natural Science Foundation of China grant 11991053. This research has made use of data obtained from the Chandra Data Archive and the Chandra Source Catalog, and software provided by the Chandra X-ray Center (CXC) in the application packages CIAO and Sherpa. This research has made use of the NASA/IPAC Extragalactic Database (NED),
which is operated by the Jet Propulsion Laboratory, California Institute of Technology,
under contract with the National Aeronautics and Space Administration. 

\section*{Data Availability}

 The Chandra data used in this paper can be downloaded from the Chandra Data Archive (PI: Brotherton; PI: Garmire). Section \ref{section3} provides detail of the data analysis and Tables \ref{tab:table1}, \ref{tab:table2}, \ref{tab:table3} \& \ref{tab:table4} lists all the measured quantities for Chandra SEAMBHs and one Archived SEAMBH. The data for other Archived SEAMBHs, super- and sub-Eddington are available in \cite{Liu+2021} (DOI: 10.3847/1538-4357/abe37f).



\bibliographystyle{mnras}
\bibliography{SEAMBH} 




\appendix
\section{}
\label{appendix}
This paper analyzed the Chandra data of four SEAMBHs that are previous studied.
Here, we compare our results with the literature values, focusing mostly on the 2-8 keV photon index as our study is based on it. We find that the our measurements are consistent with the literature values.

\begin{itemize}
    \item IRAS 04416+1215: \cite{Liu+2021} finds a photon index of $2.46_{-026}^{+0.27}$ in the $>2~ \rm keV$ Swift spectrum. A broadband XMM-Newton and NuSTAR study by \cite{Tortosa+2022} reveals that the reflected radiation dominates the primary continuum, which exhibits a slope of 1.77. \\

    \item SDSS J074352.02+271239.5: The 2-8 keV photon index $2.06^{+0.31}_{-0.30}$ was measured by \cite{Liu+2021} using Swift data. \\

    \item SDSS J100402.61+285535.3: Analysis of XMM-Newton data by \cite{Liu+2021} shows a $>2$ keV photon index of $2.31\pm0.05$, which is consistent with our measurement given the uncertainty. \\

    \item SDSSJ075051.72+245409.3: \citep{Huang+2023} finds the 0.3-8 keV  photon index of $2.22\pm0.19$ using Chandra data. Our 0.35-8 keV measurement is consistent with their findings.

\end{itemize}


\bsp	
\label{lastpage}
\end{document}